\documentclass[aps, twocolumn, notitlepage, floatfix, nofootinbib,amsmath,tighten,amssymb, showpacs]{revtex4-1} 
\usepackage{graphicx}
\usepackage{amssymb}
\usepackage{epstopdf}
\usepackage{slashed}
\usepackage{amsmath}
 \usepackage{hyperref}

\usepackage{multirow} 

\newcommand{\ignore}[1]{}
\newcommand{\be}{\begin{equation}} \newcommand{\ee}{\end{equation}}
\newcommand{\ba}{\begin{eqnarray}} \newcommand{\ea}{\end{eqnarray}}
 \renewcommand{\bf}{\textbf}

\newcommand{\GeV}{{\rm\ GeV}}

\def\slashb#1{\setbox0=\hbox{$#1$}#1\hskip-\wd0\dimen0=5pt\advance
        \dimen0 by-\ht0\advance\dimen0 by\dp0\lower0.5\dimen0\hbox
          to\wd0{\hss\sl/\/\hss}}

\input{epsf} 

\begin{document}

\title{A Theory of Ambulance Chasing}
\author{Mihailo Backovi\'{c}}
\affiliation{Center for Cosmology, Particle Physics and Phenomenology (CP3), Universit\'e catholique de Louvain, B-1348 Louvain-la-Neuve, Belgium}
\email{mihailo.backovic@uclouvain.be}
\begin{abstract}

Ambulance chasing is a common socio-scientific phenomenon in particle physics. I argue that despite the seeming complexity, it is possible to gain insight into both the qualitative and quantitative features of ambulance chasing dynamics. 
Compound-Poisson statistics suffices to accommodate the time evolution of the cumulative number of papers on a topic, where basic assumptions that the interest in the topic as well as the number of available ideas decrease with time appear to drive the time evolution. It follows that if the interest scales as an inverse power law in time, the cumulative number of papers on a topic is well described by a di-gamma function, with a distinct logarithmic behavior at large times.
In cases where the interest decreases exponentially with time, the model predicts that the total number of papers on the topic will converge to a fixed value as time goes to infinity.
I demonstrate that the two models are able to fit at least 9 specific instances of ambulance chasing in particle physics using only two free parameters. In case of the most recent ambulance chasing instance, the ATLAS $\gamma \gamma$ excess,  fits to the current data predict that the total number of papers on the topic will not exceed roughly 310 papers by the June 1. 2016, and prior to the natural cut-off for the validity of the theory. 

\end{abstract}

\maketitle

\vspace*{-13cm}
\noindent 
 CP3-16-06
\vspace*{11cm}

\section{Introduction}
In particle physics, the term ``ambulance chasing'' refers to a socio-scientific phenomenon manifest as a surge in the number of preprint papers on a particular topic. The phenomenon is usually triggered by the revelation of a new (typically speculative) experimental measurement or by a novel theory result, but necessarily before the result is experimentally confirmed as a ``discovery''. There are many examples of ambulance chasing in particle physics, the most recent of which was initiated by an announcement of a 3.5$\sigma$ anomaly in the ATLAS measurement of the di-photon spectrum around the invariant mass $m_{\gamma \gamma} = 750 \GeV $ \cite{atlasgamgam}. 

I believe it is fair to say that the motivation for engaging in ambulance chasing is mostly scientific, with a strong component of human ambition. I base the first part of the statement on a personal observation that I have yet to meet a single particle physicist who is pursuing a career in physics for any reason other than love and interest in science. \footnote{One could also come up with this conclusion by examining the prospects for long term employment in particle physics and eliminating possible alternative motives.} The component of ambition is likely a product of the unfortunate fact that much about success in particle physics depends on citation counts and $h$-indexes, where ambulance chasing serves as a mechanism for physicists to improve their bibliographic data. 

As a product of human behavior, emotion and reason, ambulance chasing is a complex system governed both by sociology and science. One would expect that as with any other dynamical system which has to account for human behavior, it would be difficult to develop a mathematical model of ambulance chasing. Yet in rare instances, dynamical systems of human behavior are driven by only a few of the many degrees of freedom and hence possible to model. In the following sections, I will argue that ambulance chasing is an example of such a dynamical system. 
Motivated by the most recent di-photon instance of ambulance chasing, I will show that it is possible to study the dynamics of ambulance chasing in a semi-analytic framework, as well as that it is possible predict the time evolution of (at least some) observables associated with ambulance chasing. Two parameter compound Poisson models are able to fit the time evolution of the cumulative number of preprint papers on an ambulance chasing topic. Furthermore, raw data appears to show that instances of ambulance chasing tend to converge to a similar time evolution at large times.

Note that the goal of this paper is not to examine the ethical aspects of ambulance chasing, but to provide some level of quantitative and qualitative understanding of the underlying dynamics. 

\section{Ambulance Chasing as a Poisson Process}

There are many observables one can construct in order to quantitatively describe the dynamics of ambulance chasing. Here I will focus on the cumulative number of papers on a topic as a function of time, $N(T)$. The reasons for choosing the observable are two-fold. First, sums tend to be more statistically stable and hence easier to study.  Second,  cumulative number of papers is an example of an observable for which it is possible to obtain real data in a reasonable amount of time. 

Let us begin with a simple assumption that on any day $t$, measured from the event which triggers the cycle of ambulance chasing, the number of papers $n$ on the topic is a random observable drawn from a Poisson distribution
$$
	P(n, t) = \frac{e^{-\mu(t) }}{n(t) !}\mu(t)^{n(t)}\,,
$$ 
where $\mu(t)$ is the mean of the distribution at time $t$.

Next, the probability that at a time $T, $ we observe $N(T) \equiv \sum_{t=1}^T n(t)$ papers is described by the compound Poisson distribution
\begin{eqnarray}
	P(N, T) &=& \prod_{t=1}^T P(n, t) \nonumber \\
		    &=& \prod_{t=1}^T \frac{e^{-\mu(t) }}{n(t) !}\mu(t)^{n(t)} \nonumber \\
		    &=& \frac{{\rm exp}\left[{-\sum_{t=1}^T \mu(t) }\right]}{N(t) !}\left[ \sum_{t=1}^T\mu(t) \right]^{N(t)},  \label{eq:compois}
\end{eqnarray}
where the last step is a result of the standard theorem of compound Poisson statistics, which states that a sum of Poisson distributed random variables is also Poisson distributed. 

Note that the expression in the last line of Eq.~\eqref{eq:compois} implies that the mean of  $P(N,T)$ is simply
\be
	\mu_N (T) \equiv \sum_{t=1}^T \mu(t). 
\ee

The functional form of $\mu(t)$ remains to be determined.  It is reasonable to assume that in ambulance chasing, both the interest in the topic and the number of available ideas which have not previously been explored decreases monotonically with time and should go to 0 as time goes to infinity.  In addition, any quantity which describes interest and the number of available ideas would by definition have to be positive definite. Hence, the behavior of $\mu(t)$ should be such that 
\ba 
\frac{d }{d t} \mu(t) \leq 0  \,\,\,\,\, 	\forall \, t, \nonumber \\
\mu(t) \geq  0 \,\,\,\,\, 	\forall \, t , \nonumber \\
  \lim_{t\rightarrow \infty} \mu(t) = 0. \label{eq:conditions}
\ea

This leads to an ansatz that $\mu(t)$ can quite generically be written as
\be
	\mu(t) =   
	\left\{
\begin{array}{cr} 
      \sum_{k=1}^\infty \frac{a_k}{t^k} & {\rm \,\, model\, 1}\\
      A\, {\rm exp}\left(-\sum_{k=1}^\infty B_k\, t^k\right) & {\rm \,\,model \, 2}
    \end{array} \right. ,
    \label{eq:mu}
\ee
where $a_k, A, B$ are the constants in time. The generic expansion of inverse powers in $t$ (model 1) encompasses more general functional forms, ($e. g.$ $\mu(t) \sim {\rm tan}^{-1}(t)$), while model 2 serves to describe cases where $\mu(t)$ might decrease faster than any power of $1/t$. Other than the expressions in Eq.~\eqref{eq:mu} I can not think of any other case which satisfies the conditions of  Eq.~\eqref{eq:conditions} that is not already well approximated by models 1 and 2.

The ansatz is surely not perfect as one could easily imagine situations where the appearance of an ambulance chasing paper will actually induce interest in the community and hence result in additional publications, manifest for instance as terms proportional to $t^n$, where $n\geq 0$ in model 1. However, these instances tend not to last long and result in fluctuations which can be absorbed into the coefficients. 

Let us first examine model 1 in more detail. 
As $t\rightarrow \infty, $ the dominant term in the $\mu(t)$ expansion is  $\sim 1/t$, assuming that there is no large hierarchy between the coefficients  (which should be the case in any natural theory). 

At late times, we can hence drop terms with $k > 1$, leading to 
\be
	\mu_N(T) =\sum_{t=1}^T\frac{a}{t} = a\, H (T), \label{eq:muNT}
\ee
where $H(T)$ is a harmonic number of $T$. 

As harmonic numbers of large arguments scale logarithmically, the immediate implication of Eq.~\eqref{eq:muNT} is that the distribution of $N(T)$  asymptotically approaches a Poisson distribution with the mean
$$ \lim_{T\rightarrow \infty} \mu_N(T) =c_1 + c_2\, {\rm log}(T) ,$$
where $c_{1,2}$ are constants. The logarithmic divergence of the model is not an issue, as in each ambulance chasing instance, there exists a cut-off time beyond which the model is invalid. The cut-off is typically determined by the time at which the result is either confirmed or refuted, in which case the assumptions behind the motivation for $\mu(t) \sim 1/t$ do not hold anymore. 

In the above derivation I assumed that time between two successive data points $n(t)$ flows in uniform discrete steps. This is not strictly true in practice as the preprint publication dates are skewed by weekends and holidays. In order to mitigate this effect, I will introduce another parameter into the definition of $\mu_N(T)$ as
\be
	\mu_N(T) = a\, H(b\, T). \label{eq:moneyeq}
\ee
The $b$ parameter also helps to capture the effect of higher $k$ terms in the expansion, which can improve the fit at smaller $T$. 
Note that  introducing $b$ into the definition preserves the characteristic logarithmic form of $\mu_N(T)$ as $T\rightarrow \infty$. The $b$ parameter also  analytically continues the argument of the Harmonic number into the real plane. This is not a problem, as Harmonic numbers analytically continue into the real plane via the di-gamma function $\psi(x)$. In the following, I will continue to use the symbol $H(x)$, and implicitly assume the analytic continuation.  

An analogous calculation using model 2, again keeping only the leading term $\sim e^{- t},$ leads to an expression
\be
	\mu_N(T) =  A' \,\left[ 1-e^{-BT }  \right], \,\,\, A' \equiv \frac{A}{e^{B\cdot {\rm day}} - 1}      .\label{eq:moneyeq2}
\ee

The behavior of $\mu_N(T)$ in case of model 2 is quite different in the limit of $T\rightarrow \infty$, where one finds
$$
	 \lim_{T\rightarrow \infty} \mu_N(T) =  A'= {\rm const.} \, ,
$$
implying that model 2 generically predicts lower values of $N(T)$ as $T \rightarrow \infty$. This is consistent with the model 2 assumption that the interest and available number of topics will decrease with time faster than any power of $1/t$. 

\begin{figure*}[!]
\includegraphics[width=2.2 in]{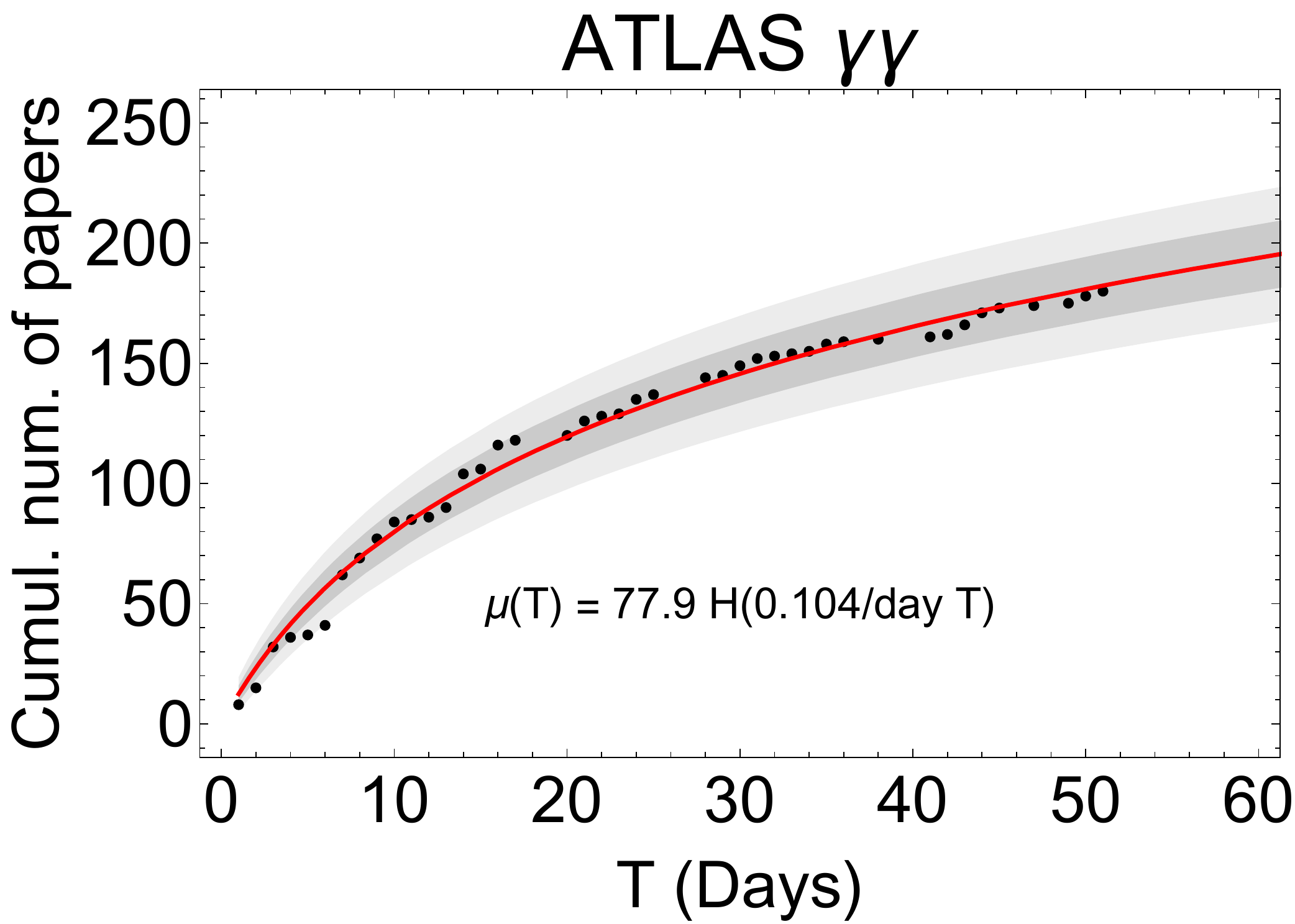}
\includegraphics[width=2.2 in]{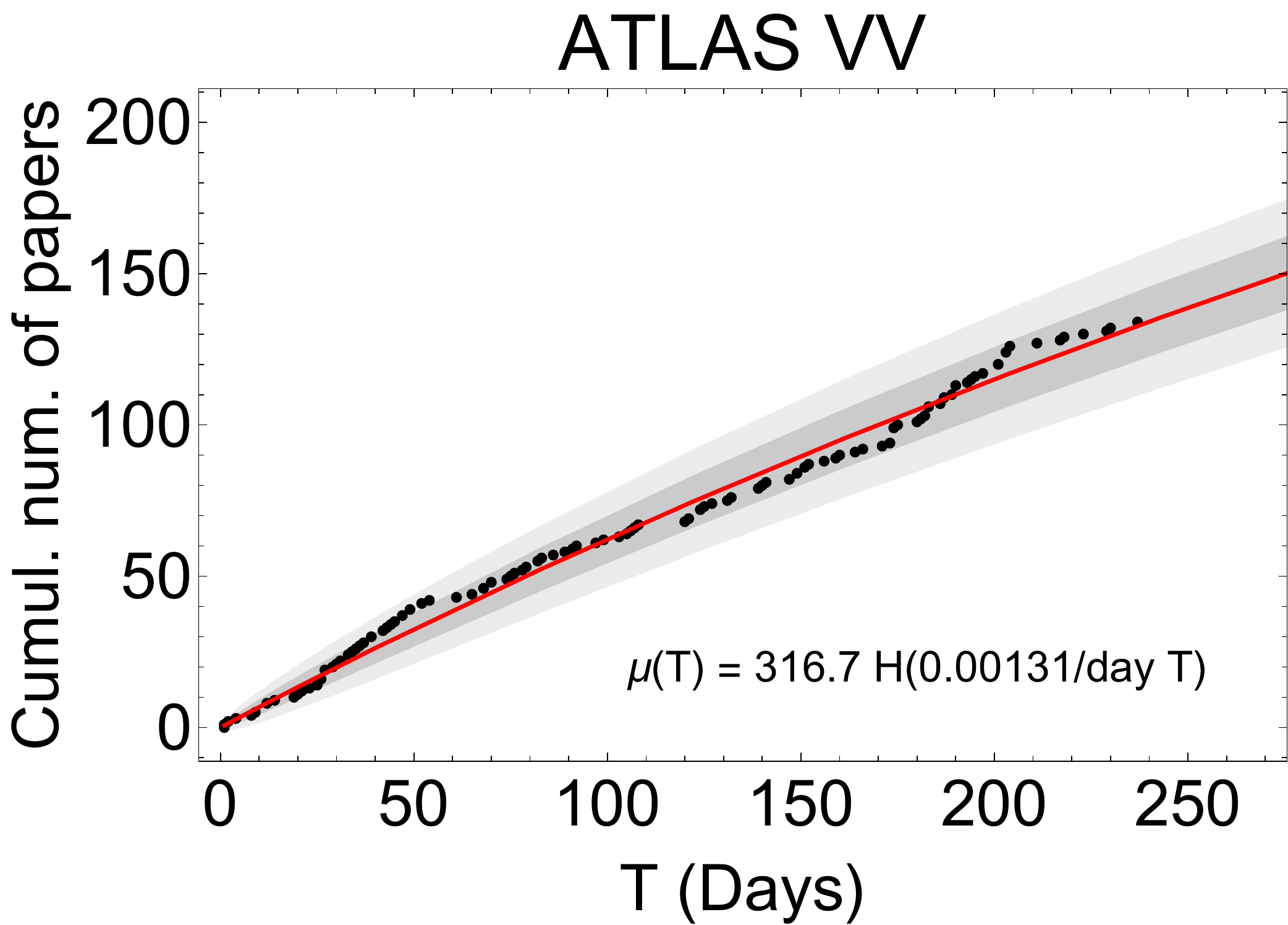}
\includegraphics[width=2.3 in]{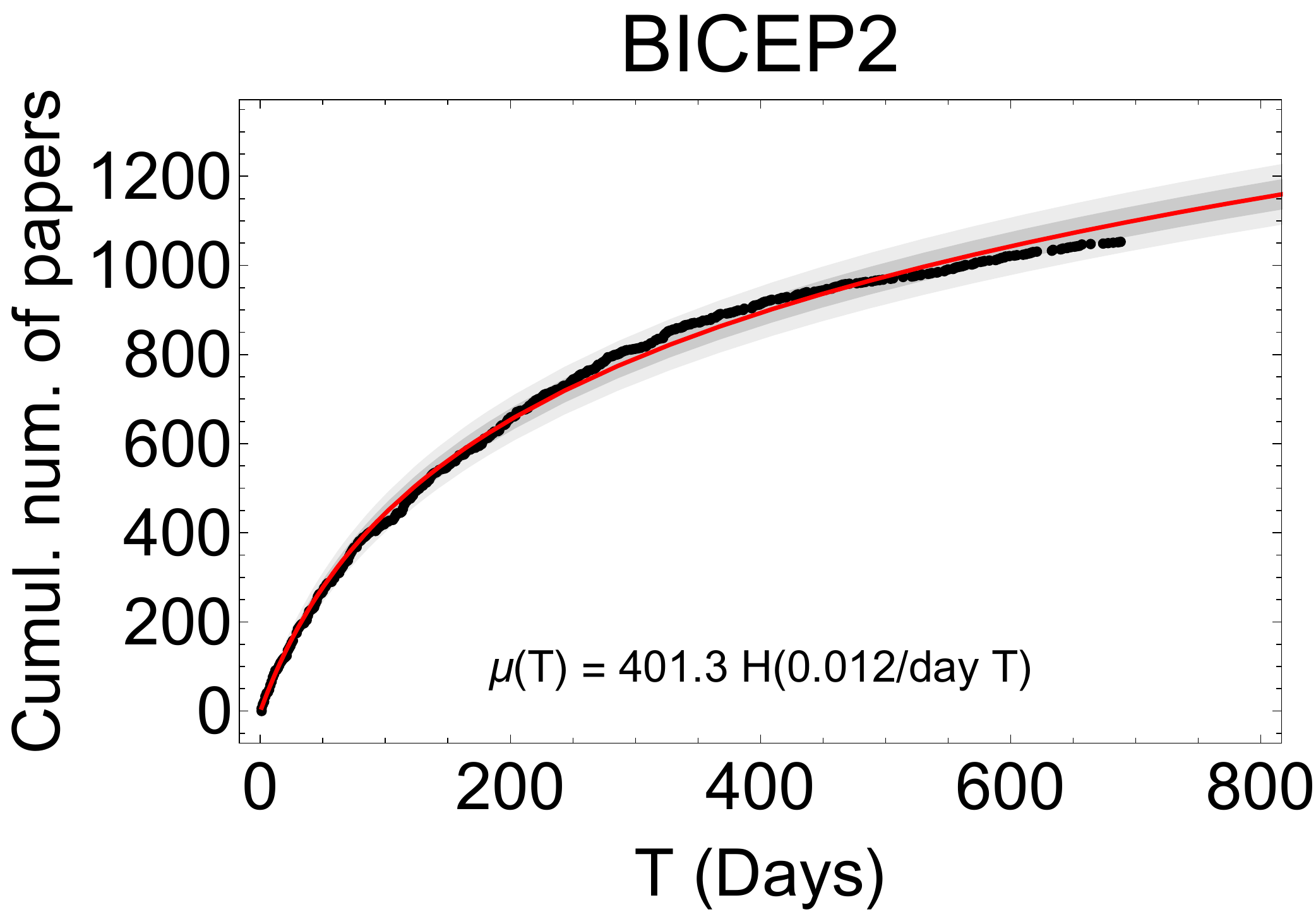}\\
\includegraphics[width=2.2 in]{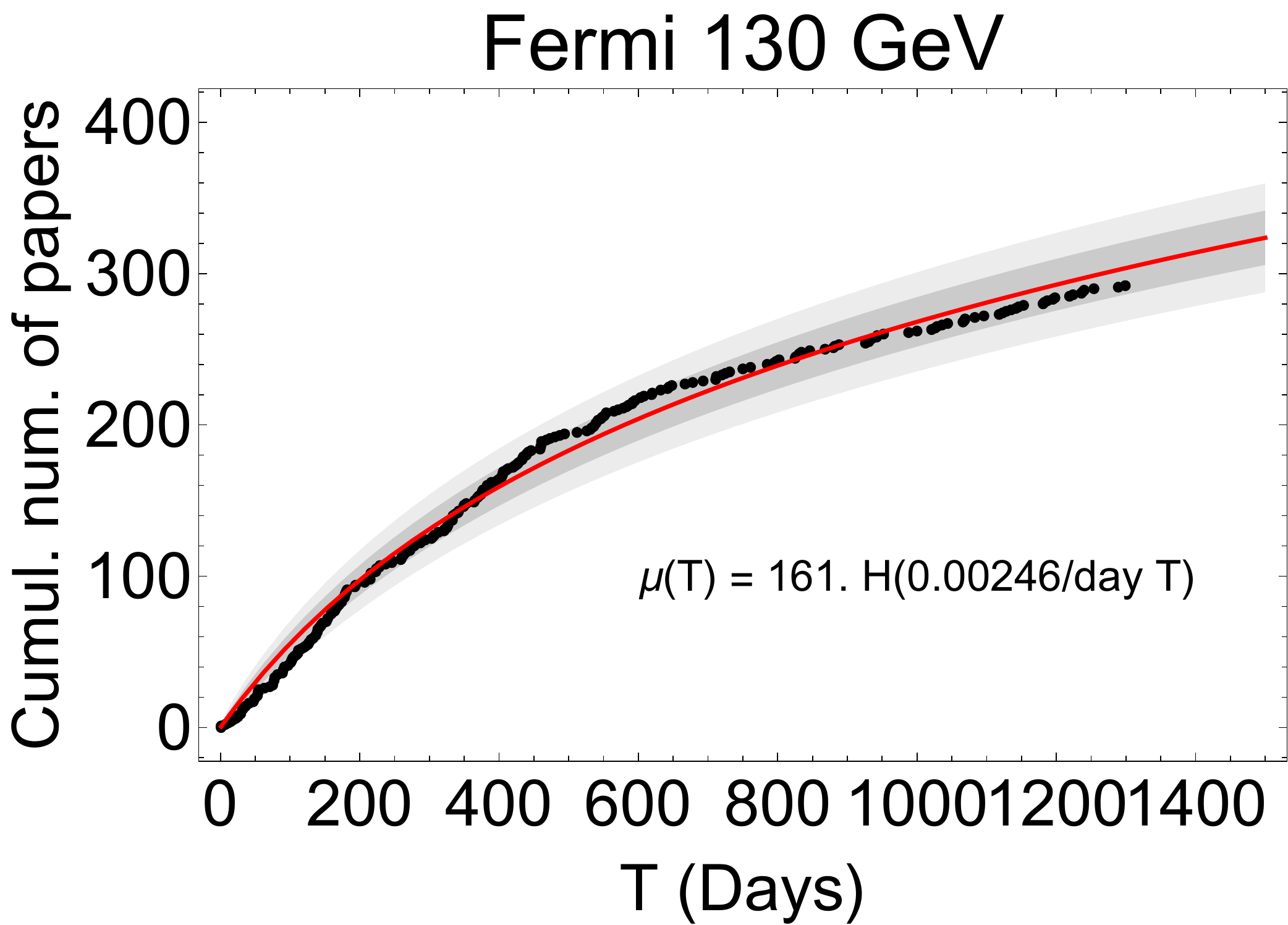}
\includegraphics[width=2.2 in]{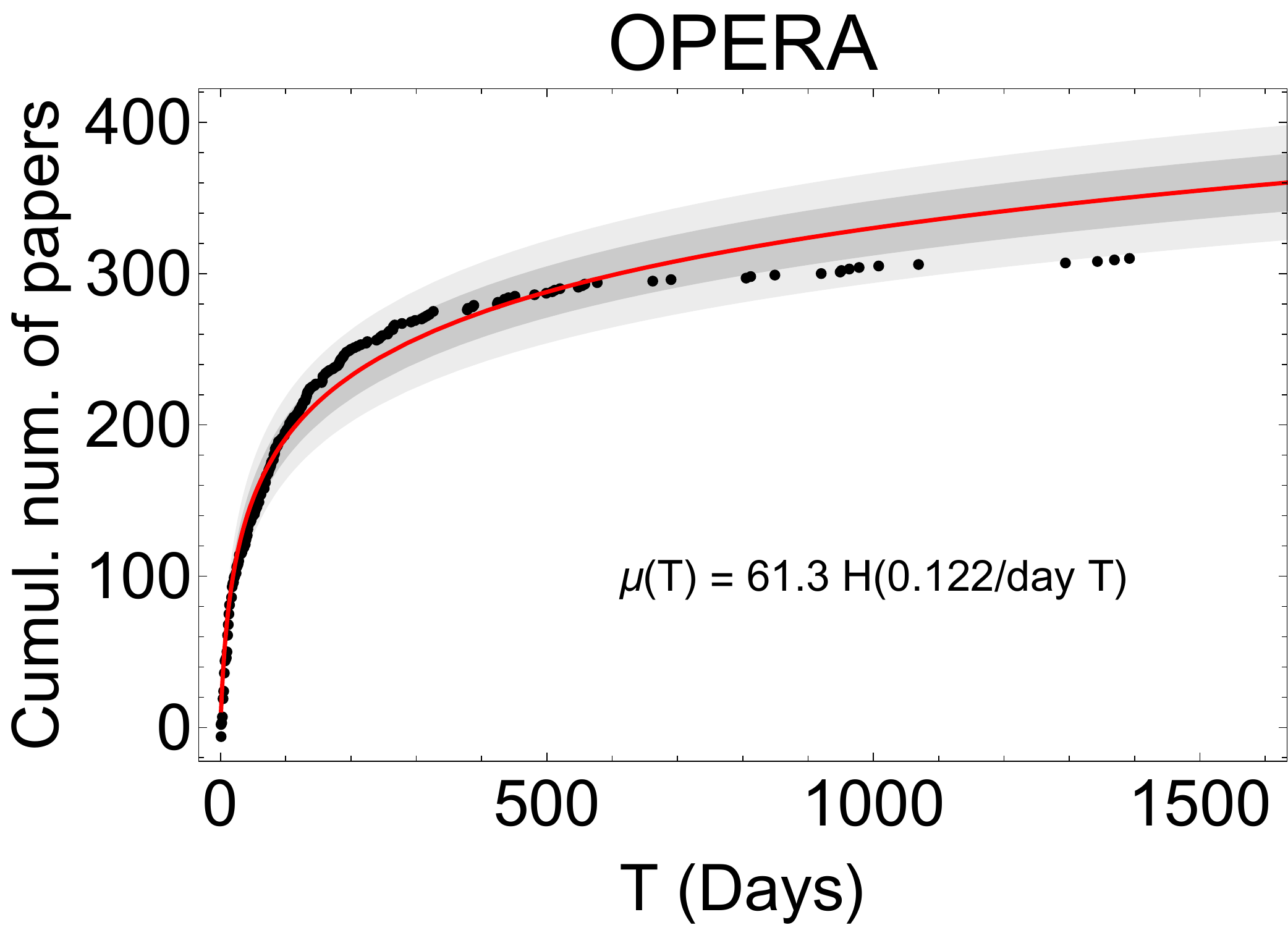}
\includegraphics[width=2.2 in]{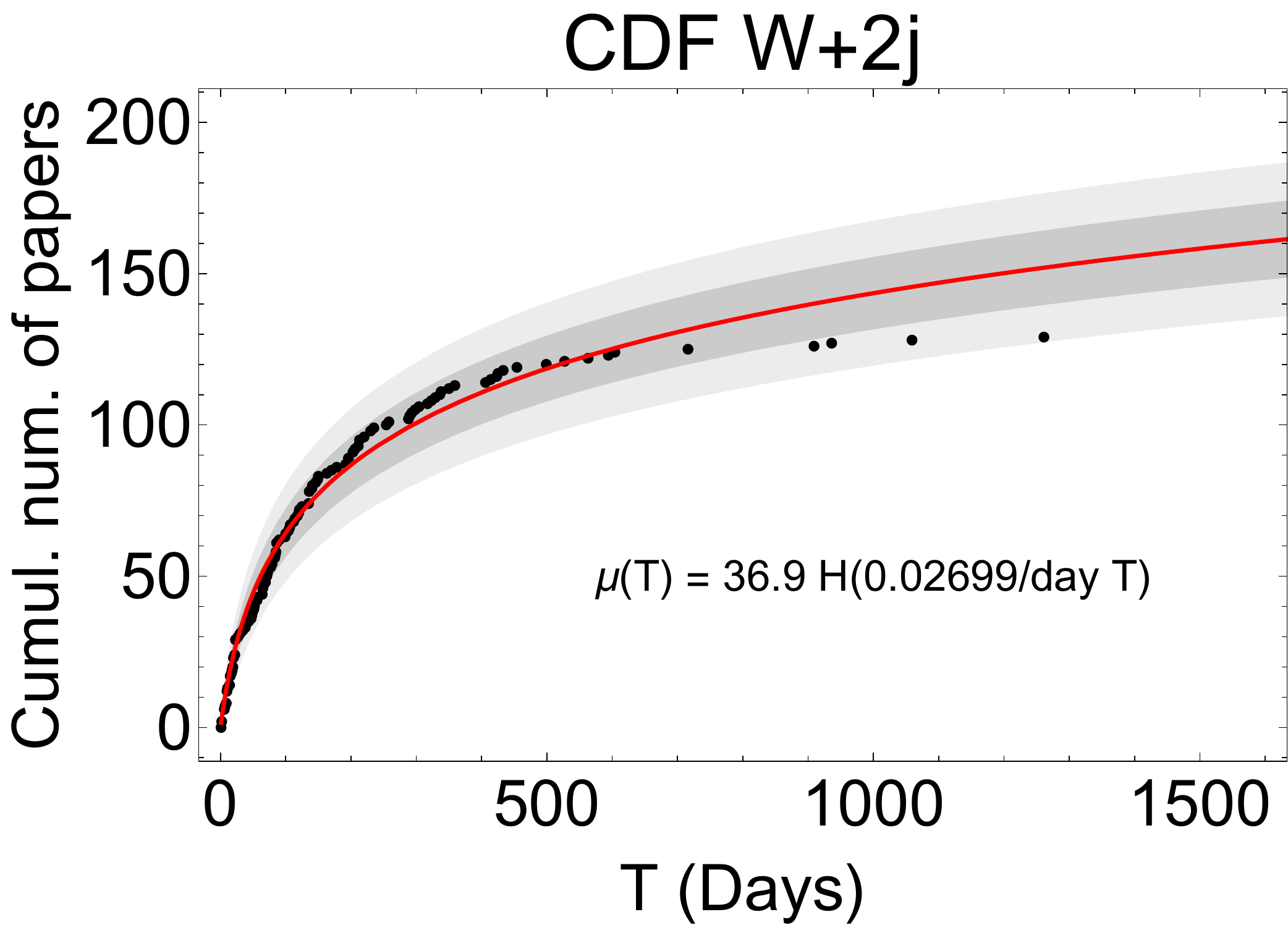} \\
\includegraphics[width=2.3 in]{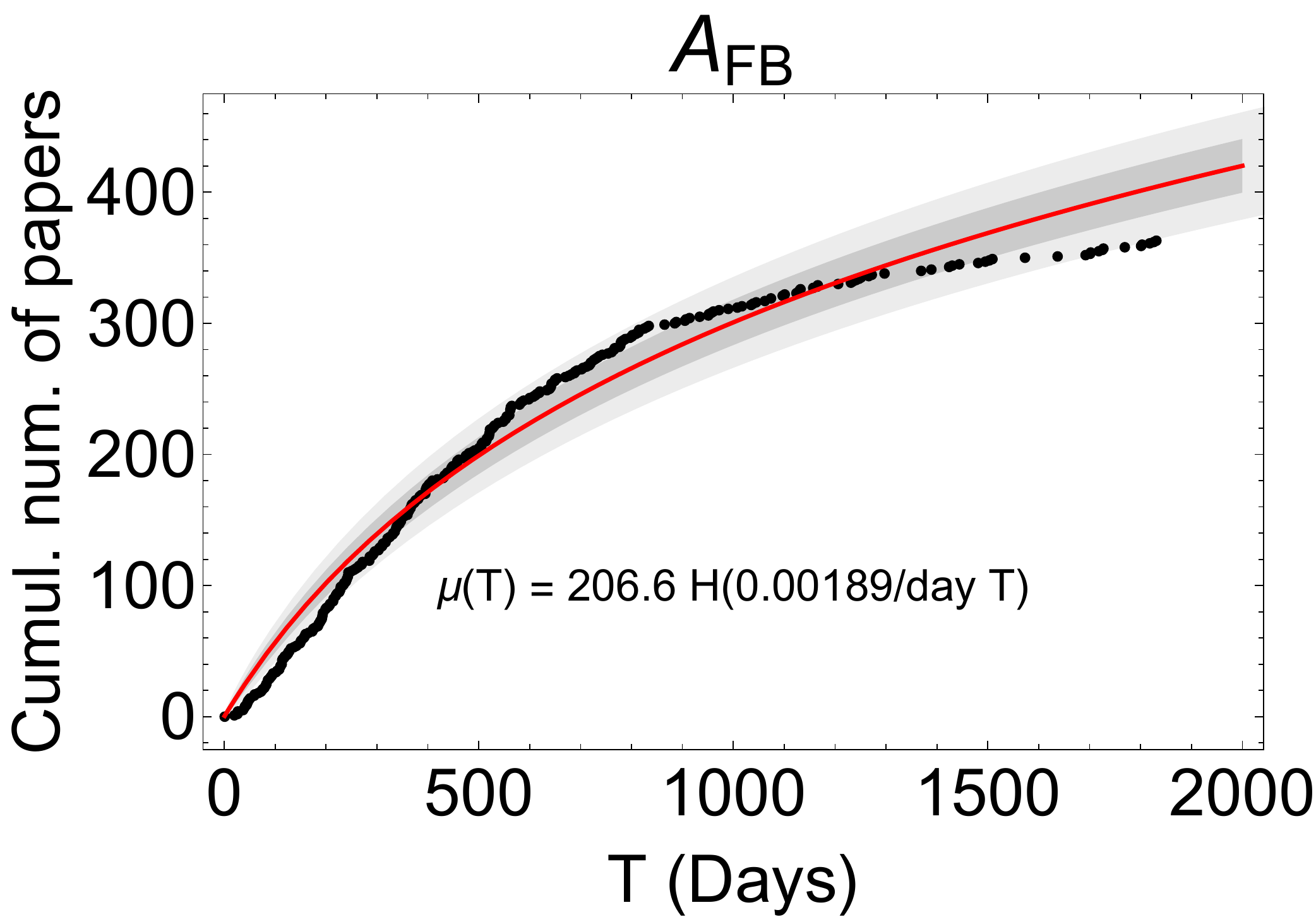}
\includegraphics[width=2.3 in]{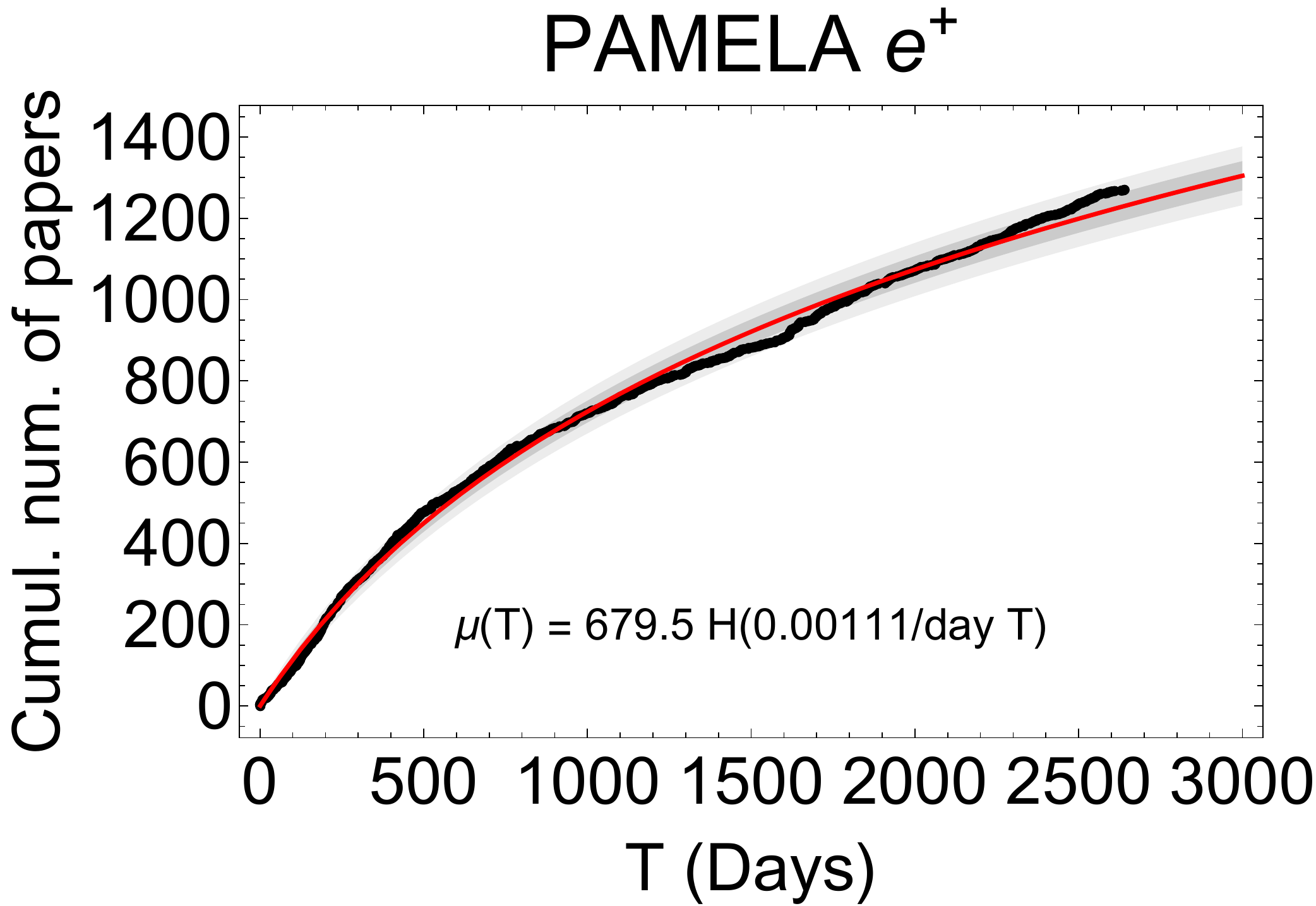}
\includegraphics[width=2.2 in]{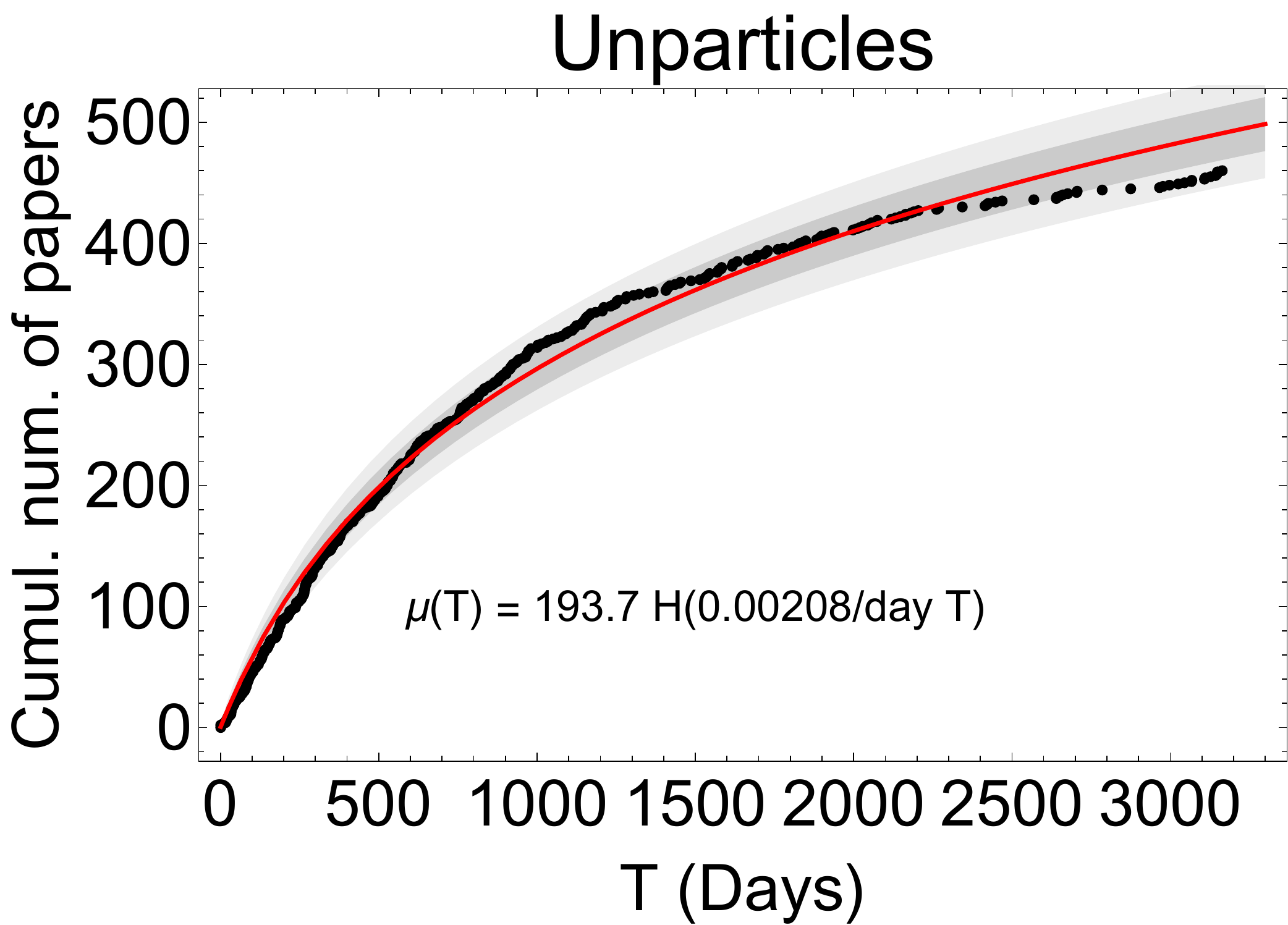}

\caption{Cumulative number of papers on a particle physics topic as a function of time, using \textbf{model 1}. The points show raw data extracted from the citation counts of papers in Table~\ref{tab:data}. The red line is the fit for the Poisson mean $\mu(T)$ in Eq.~\eqref{eq:moneyeq}. The bands represent the one and two Poisson standard deviations, i.e. $\sqrt{\mu(T)}$.}\label{fig:moneyplot1} 
\end{figure*}

\section{Results}

In order to test the models from the previous section, I extracted the data on several recent instances of ambulance chasing from the \verb|inSPIRE| and \verb|arXiv| repositories, where I obtained the cumulative number of papers on a topic as a function of time, $N(T)$, by extracting lists of citations to the result which initiated the ambulance chasing instance. The method is not perfect, as a number of papers that are not closely related to the topic will still cite the experimental result.
In addition, I only considered papers which had an \verb|arXiv| number assigned, as it was otherwise difficult to determine the date of the publication in an automated fashion. 
I expect that these approximations will result in  systematic errors of $~O(10) $ papers in total (per data set) and will hence typically be smaller than the statistical error.

\begin{table}[t]
\footnotesize
\begin{center}
\begin{tabular}{lcc}
\hline
     Result & Announcement Date & \verb|arXiv| number \\
\hline
    ATLAS $\gamma \gamma$  \cite{atlasgamgam} &15 Dec. 2015   & N/A\\
    ATLAS $V V$ \cite{Aad:2015owa} & 2 Jun.  2015& 1506.00962  \\
    BICEP2 \cite{Ade:2014xna} & 17 Mar. 2014& 1403.3985   \\
    Fermi 130 GeV \cite{Weniger:2012tx} & 12 Apr. 2012 &  1204.2797 \\
    OPERA \cite{Adam:2011faa} &22 Sep. 2011 & 1109.4897 \\
    CDF $W+2j$ \cite{Aaltonen:2011mk} & 4 Apr. 2011 &1104.0699 \\
    $A_{FB}$ \cite{Aaltonen:2011kc}& 30 Dec. 2010 &  1101.0034 \\
    PAMELA $e+$\cite{Adriani:2008zr} & 8 Oct. 2008 &  0810.4995 \\
    Unparticles \cite{Georgi:2007ek} & 23 Mar. 2007& 0703260  \\
\hline

\end{tabular}
\caption{Recent instances of ambulance chasing in particle physics.}
\end{center}
\label{tab:data}
\end{table}

\begin{figure*}[t]
\includegraphics[width=2.2 in]{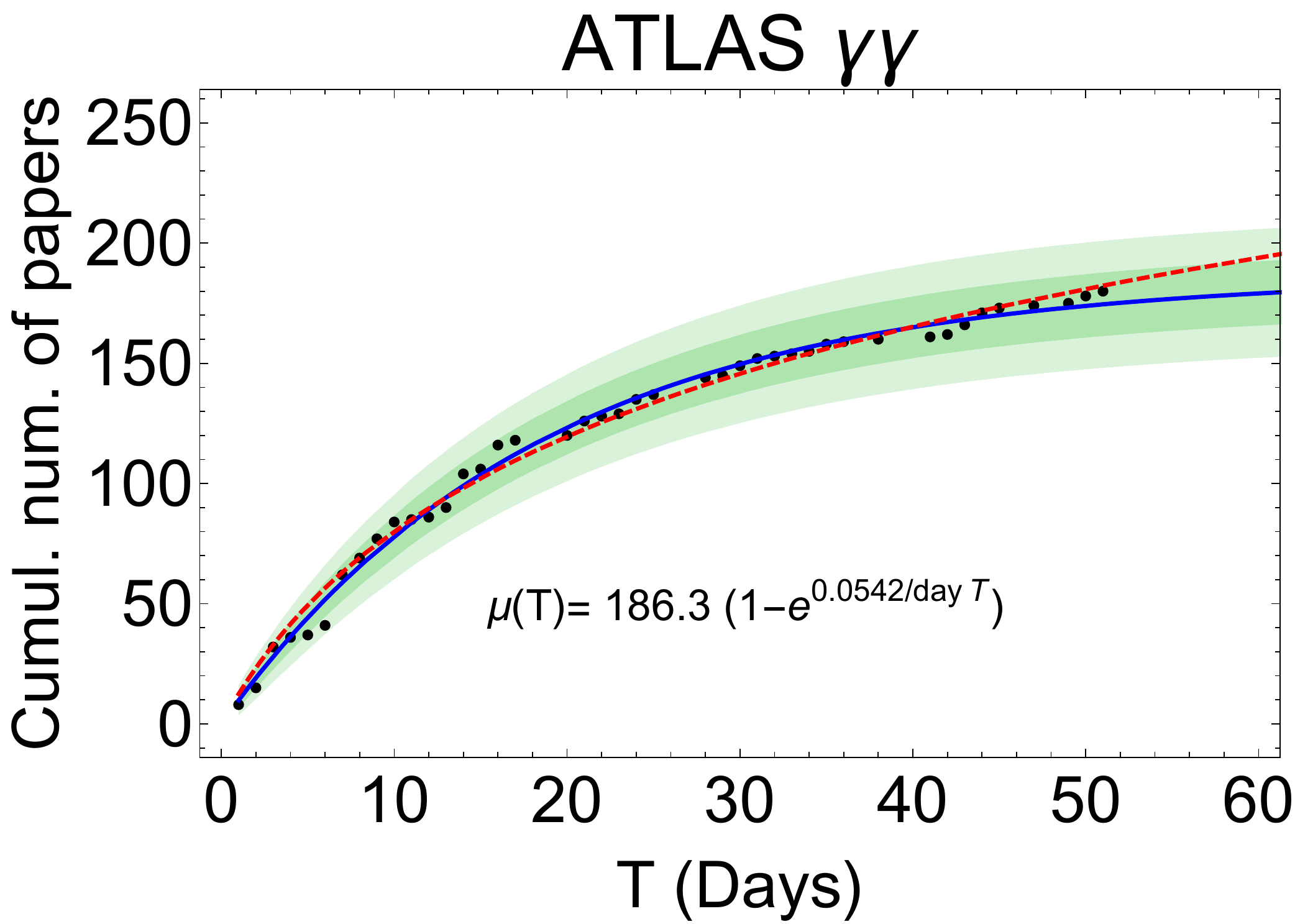}
\includegraphics[width=2.2 in]{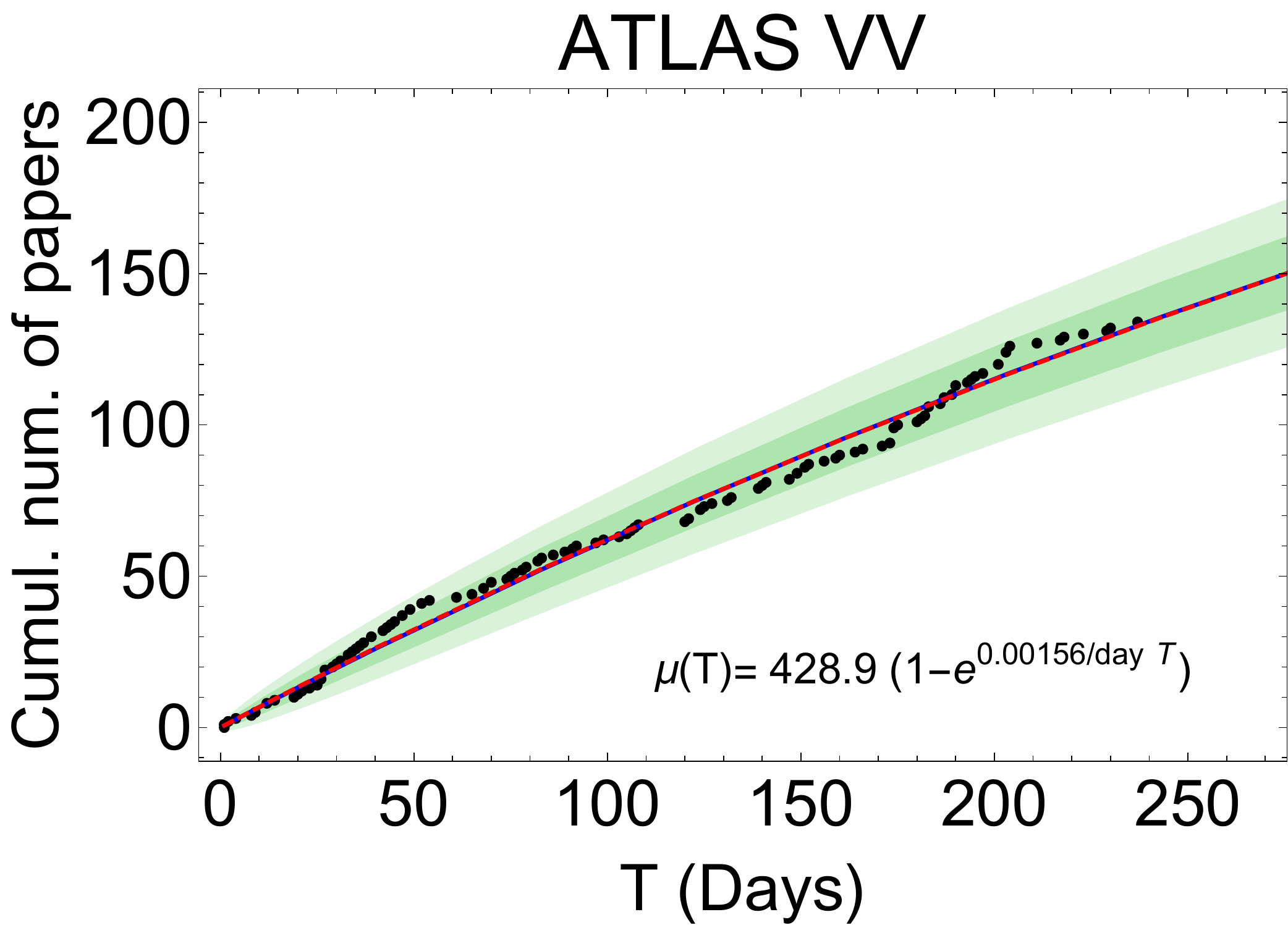}
\includegraphics[width=2.3 in]{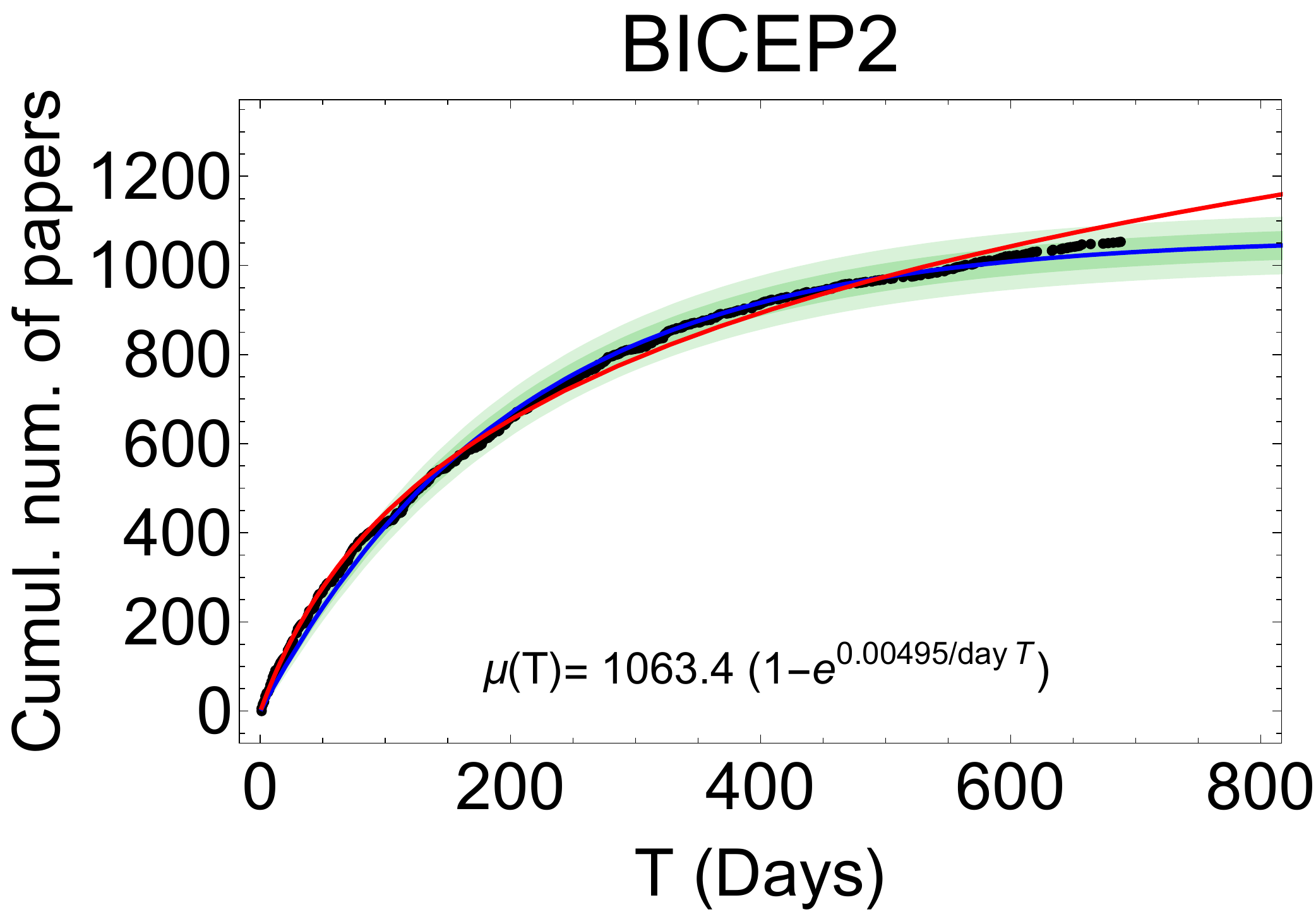}\\
\includegraphics[width=2.2 in]{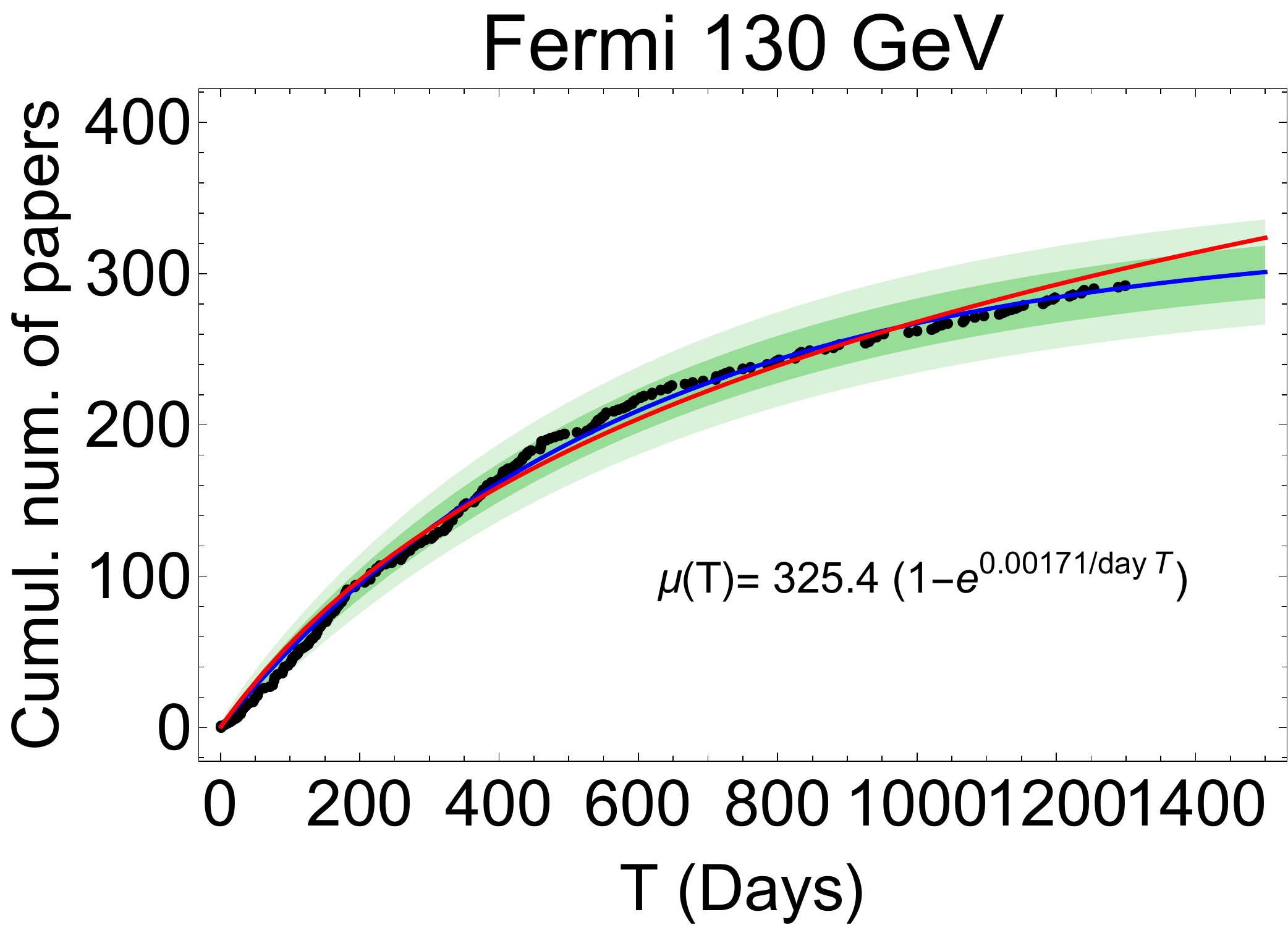}
\includegraphics[width=2.2 in]{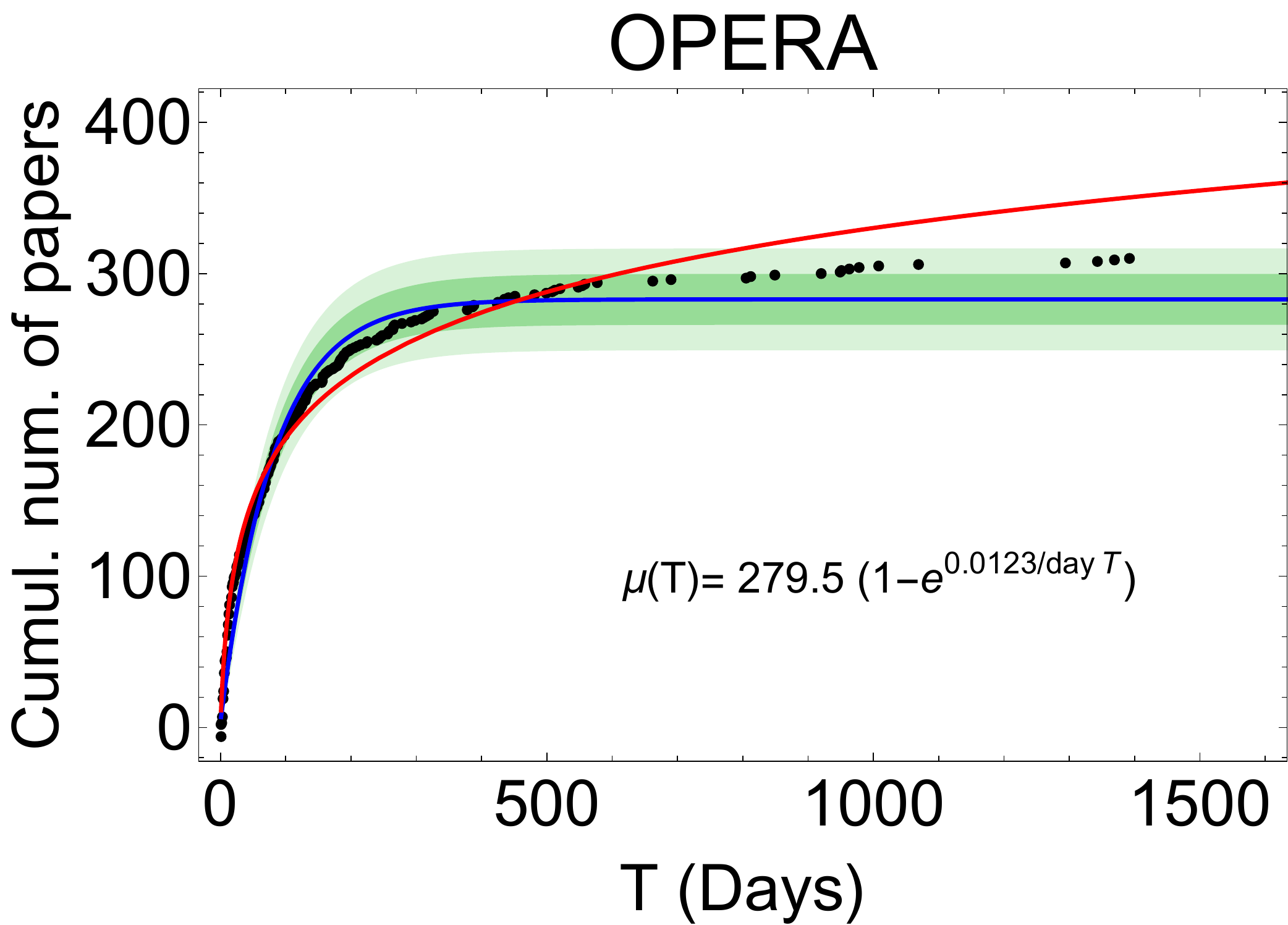}
\includegraphics[width=2.2 in]{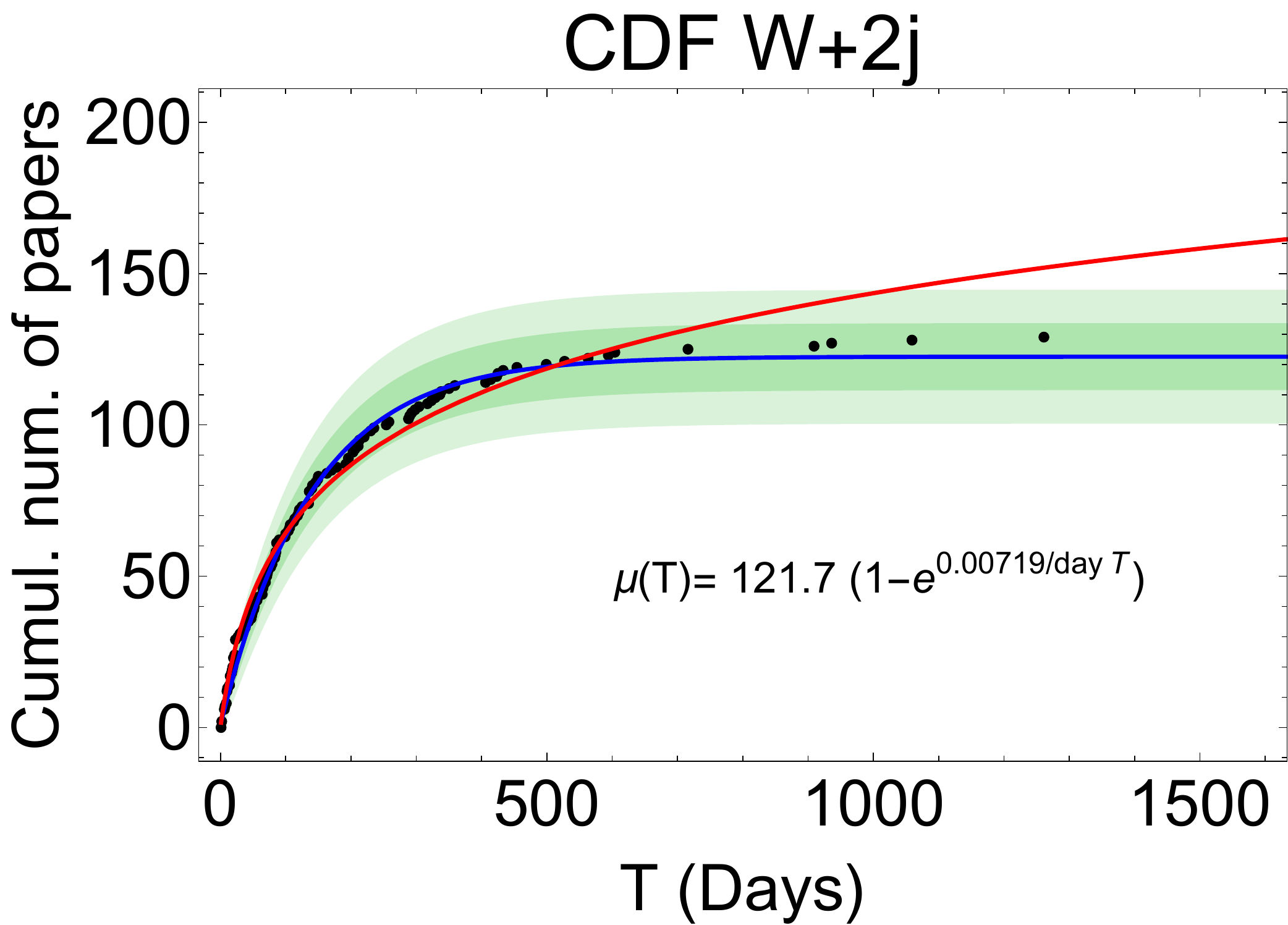} \\
\includegraphics[width=2.3 in]{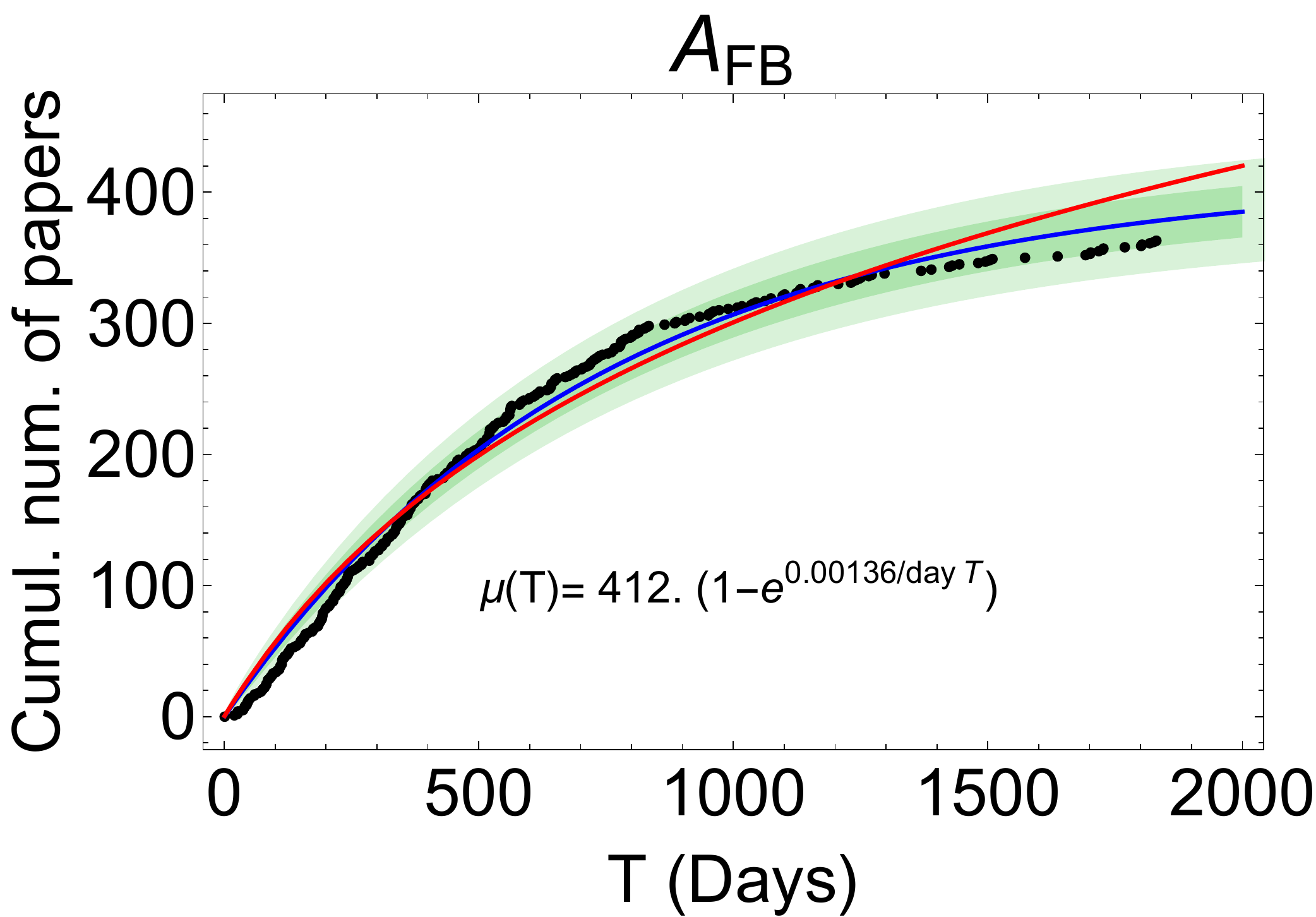}
\includegraphics[width=2.3 in]{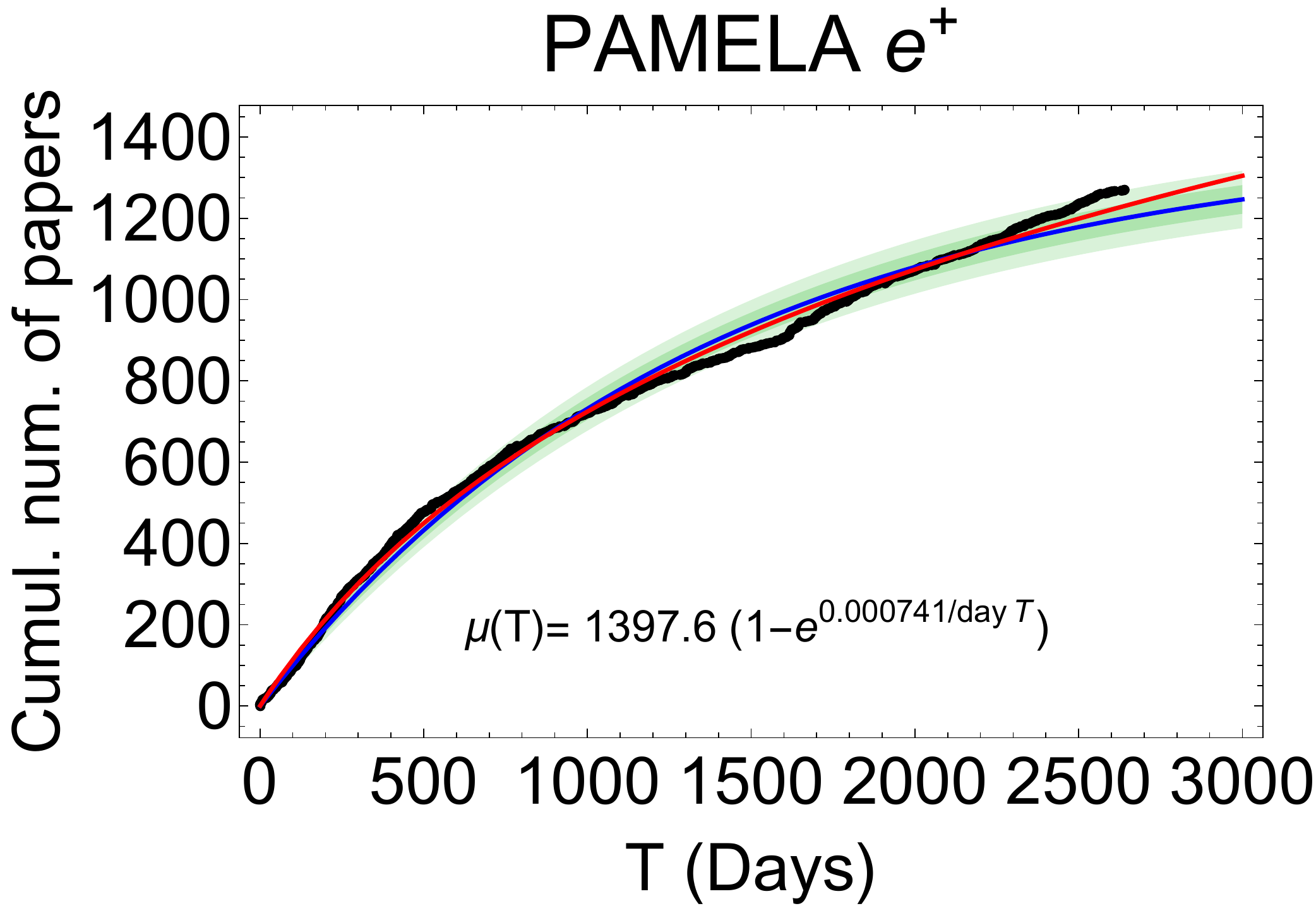}
\includegraphics[width=2.2 in]{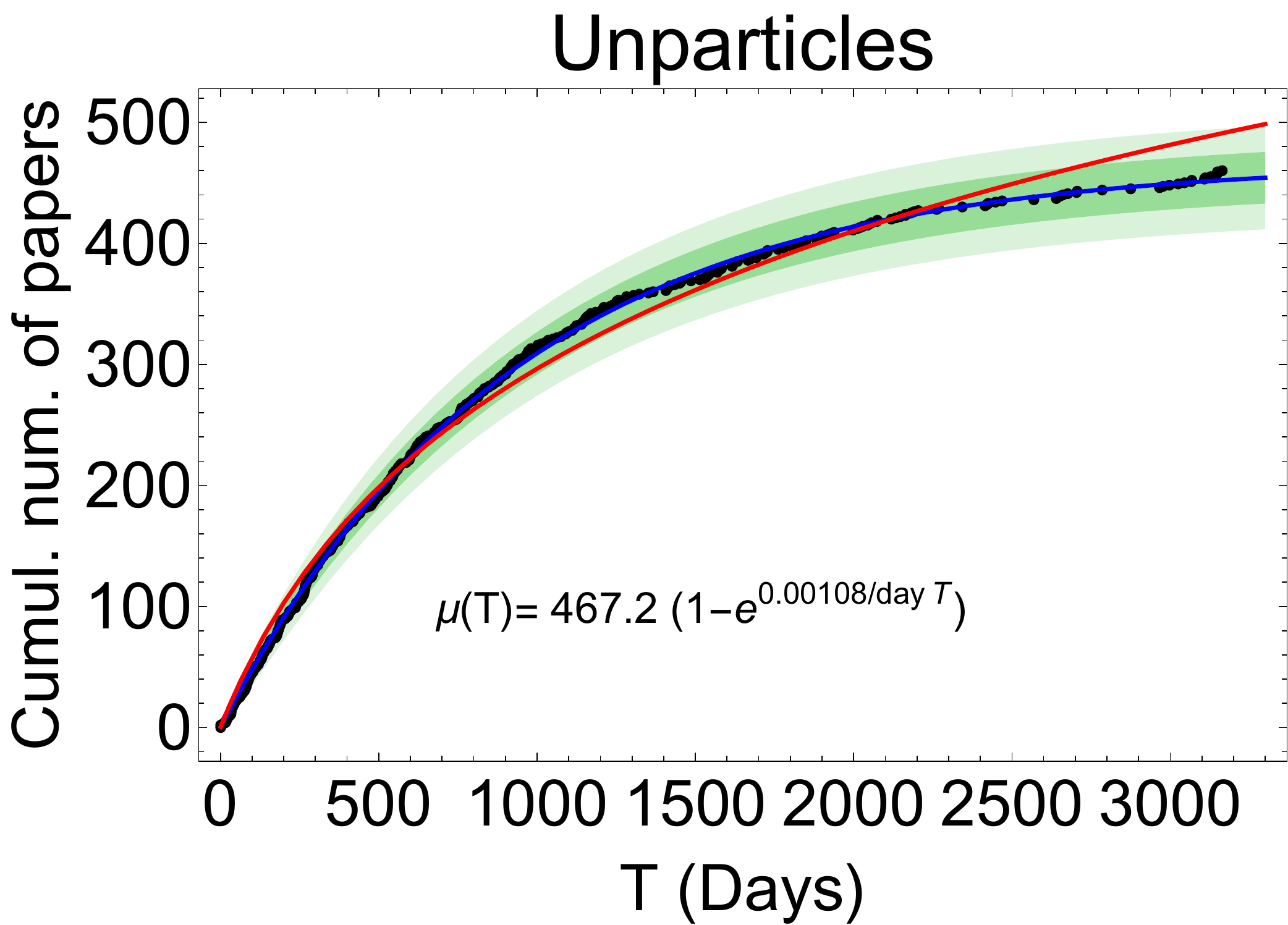}

\caption{Cumulative number of papers on a particle physics topic as a function of time, using \textbf{model 2}. The points show raw data extracted from the citation counts of papers in Table~\ref{tab:data}. The red line is the fit for the Poisson mean $\mu(T)$ in Eq.~\eqref{eq:moneyeq}. The bands represent the one and two Poisson standard deviations, i.e. $\sqrt{\mu(T)}$. The red lines represent the best fit for $\mu(T)$ using model 1 for reference.}\label{fig:moneyplot2} 
\end{figure*}

For each data set, the date on which the result responsible for the cycle of ambulance chasing is announced can be established as the ``zero time.'' The choice is not unique as one could easily repeat the exercise by considering, say, the date of the first preprint as the starting date in each data set. Note, however, that none of the general conclusions in this paper are affected by the choice of the reference date, as the choice of  zero only shifts the data sets and does not affect any functional forms.

\begin{figure*}[t]
\includegraphics[width=3.1 in]{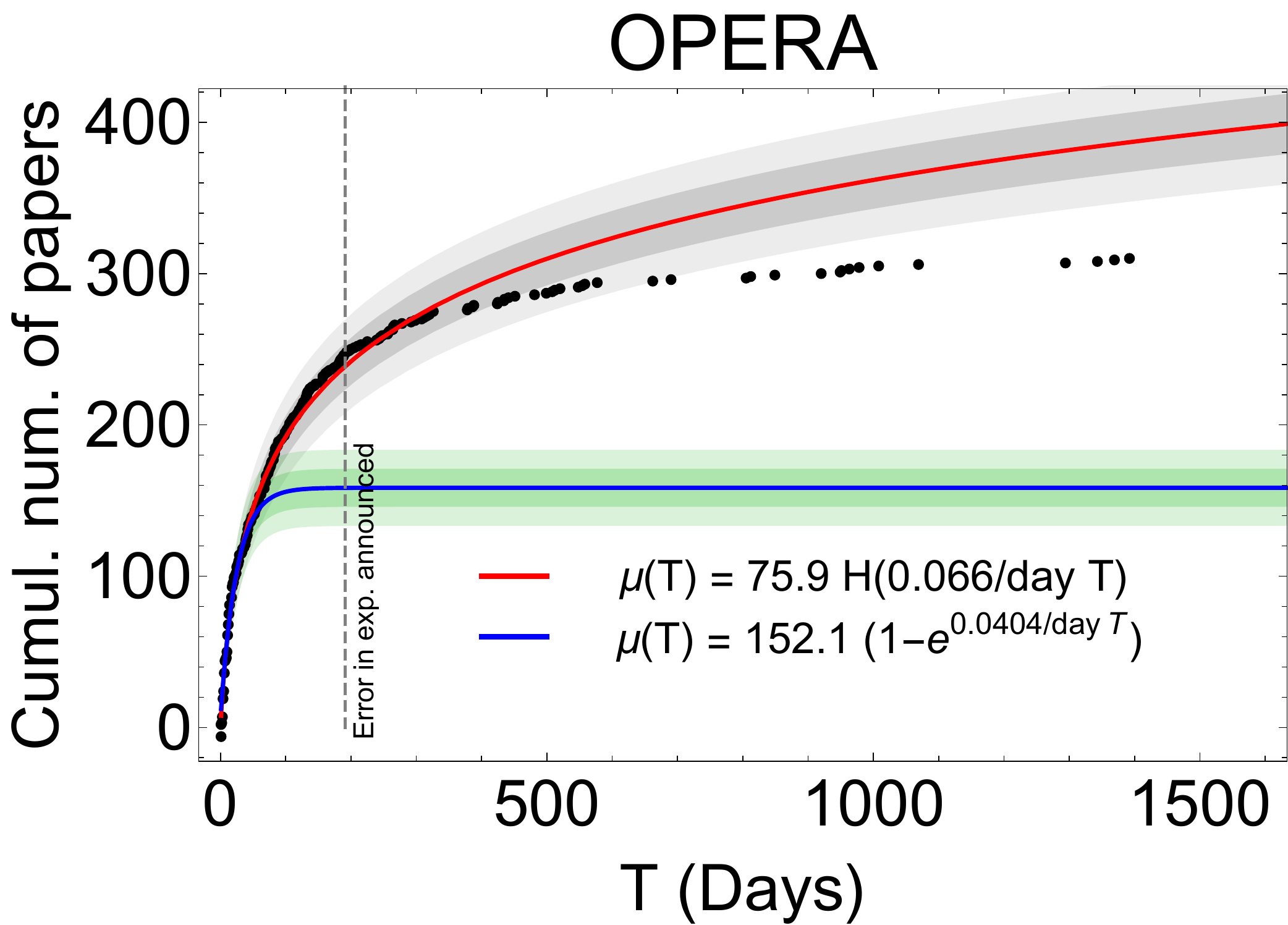} \vspace{0.1in}
\includegraphics[width=3.3 in]{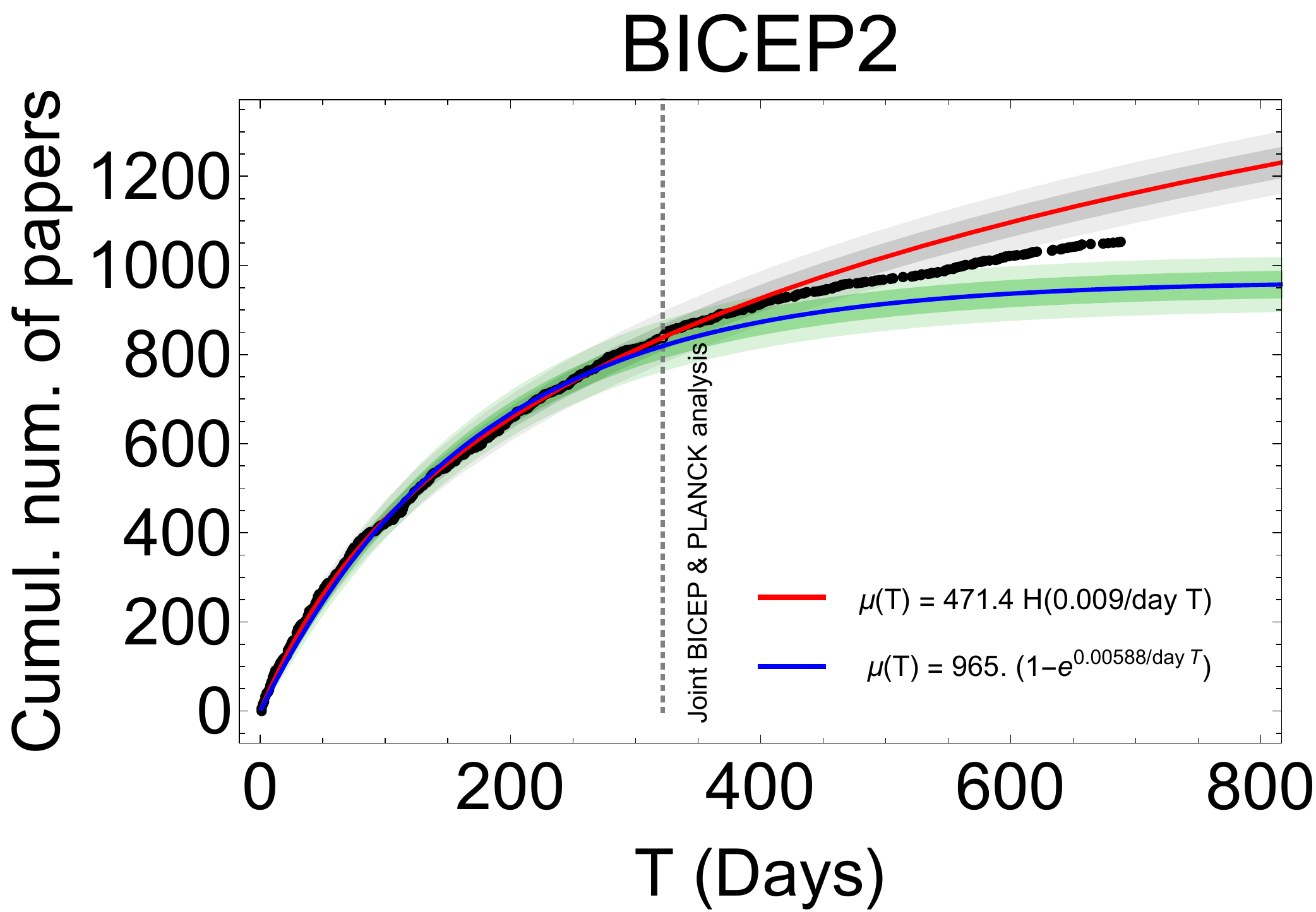}
\caption{Fit to OPERA and the BICEP 2 data, using only the data up to the announcement of the experimental error.}\label{fig:opera} 
\end{figure*}

Table \ref{tab:data} shows a summary of the data sets considered in this paper, together with the dates of appearance as well as the \verb|arXiv| number of the original note whose citation data I use as an estimate of $N(T)$.

A compound Poisson distribution with a mean of Eqns.~\eqref{eq:moneyeq}  and ~\eqref{eq:moneyeq2} is able to fit each of the ambulance chasing instances in Table \ref{tab:data}, within the $2\sigma$ Poisson bands. 
Fig. \ref{fig:moneyplot1} shows the results for model 1, where the red lines show the fit of the mean in Eq.~\eqref{eq:moneyeq} to the data and the gray bands show one and two Poisson standard deviations. In each case, I find that the general functional form of Eq.~\eqref{eq:moneyeq} fits the time evolution of $N(T)$ well, and that the data rarely exceeds the $2\sigma$ bands from the fit. In cases such as OPERA, BICEP 2 and CDF $W+2j$, where the result which initiated the cycle of ambulance chasing was eventually refuted, there should be no expectation that the model of Eq.~\eqref{eq:moneyeq} fit the data well at later times. This is simply due to the fact that after the result is refuted, the cut-off for the validity of the theory is clearly reached. Still, the $b$ parameter in Eq.~\eqref{eq:moneyeq} is able to mitigate some of this effect and still provide a satisfactory fit to the data even after the cut-off is reached.

 Fig.~\ref{fig:moneyplot2} shows the results of model 2 fits, where I show the fit to $\mu_N(T)$ from model 1 as the red line for reference. In all instances, model 2 appears to fit the \textit{overall} data equally well at small $T$, while the fit obtained with model 2 is often different, and superior at large $T$. Perhaps the most striking result using model 2 is the quality of fit in the case of Unparticles, where the data shows almost no deviation from $\mu_N(T)$ using model 2. This is strongly suggestive of the initially high enthusiasm about Unparticles which was followed by an exponentially decaying interest in the topic at later times. 

The reason that model 2 in many cases fits the overall data better than model 1 is likely due to the fact that model 1 fits in Fig.~\ref{fig:moneyplot1} are performed including the data beyond the cut-off for the model validity, while model 2 is able to accommodate for the cut-off in a more natural fashion, as we expect $N(T) \approx {\rm const}$ after the cut-off has been reached. A more fair comparison between the two models would have to be done using only the data up to the cut-off, but it is often non-trivial to determine a precise date at which the theory breaks down. 

The OPERA result is perhaps the best example of a system with a clearly defined cut-off, where the revelation of an experimental error roughly 190 days after the original announcement represents a clear cut-off for the validity of the theory. Fitting the OPERA data up until the cut-off date provides a better fit to the early time data in model 1, but is naturally much  worse after the cut-off has been reached, as shown in Fig. \ref{fig:opera}. Conversely, fitting model 2 only to the data before the cut-off results in a significantly worse fit compared to model 1. Another example is the recent BICEP 2 result, where roughly 320 days after the announcement, a joint PLANCK-BICEP analysis of the cosmic dust foreground confirmed the long suspected over-estimate of the BICEP 2 signal. A fit to the BICEP 2 data using only the points before the joint announcement is also shown in Fig. \ref{fig:opera}.

\begin{figure}[t]
\includegraphics[width=3.2in]{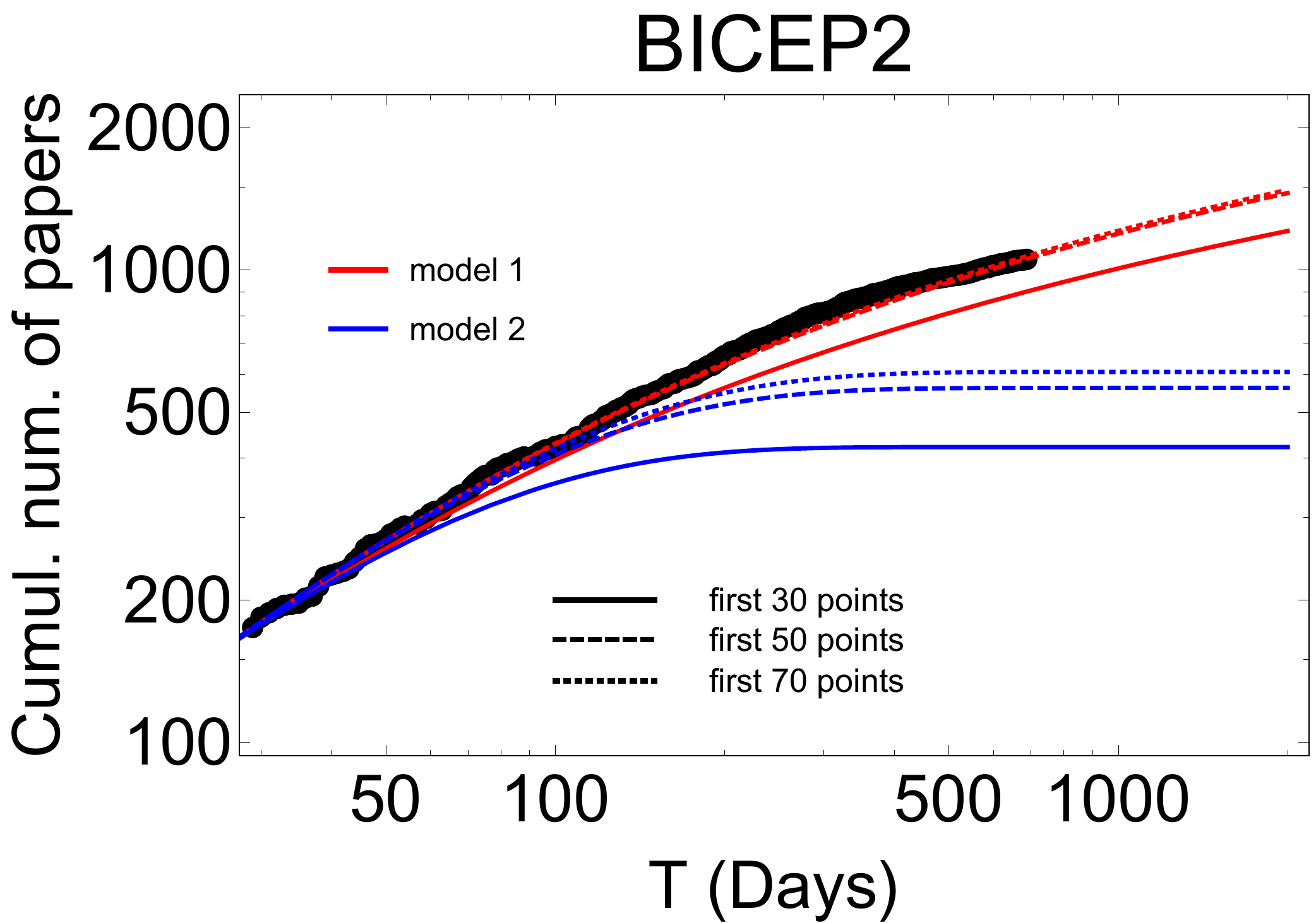}
\caption{Forecasting abilities of models 1 and 2. The blue(red) lines are associated with fits to model 1(2) respectively, while the solid, dashed and dotted lines represent best fits using the first 30, 50 and 70 data points respectively. }\label{fig:fitsm1m2} 
\end{figure}

Fig. \ref{fig:opera} reveals an important point about the differences in models 1 and 2. While model 2 appears to be able to fit the overall data better, the ability of model 2 to forecast the time evolution of $N(T)$ appears inferior to model 1. The reason is that the exponential suppression of $\mu(t)$ characteristic of model 2 typically forces $\mu_N(T)$ to a constant soon after the range of the fit data is exceeded. This can be seen better in Fig.~\ref{fig:fitsm1m2} where I used the BICEP 2 data as an illustration. The solid, dashed and dotted curves show $\mu_N(T)$ fits to the BICEP~2 data using the first 30,50 and 70 data points respectively, where model 1 is represented in red, while model 2 is represented in blue. Fig.~\ref{fig:fitsm1m2} suggests that the fit to model 2 is much more sensitive to the amount of data used in the fit, where in each instance, the exponential suppression seems to take over soon after the data used in the fit is exceeded in time. This occurs because the limit as $T\rightarrow \infty$ of model 2 is set to a constant by the value of the $A'$ parameter. Since the functional form of model 2 is monotonically increasing, the model can hence only forecast the data up to $N(T) = A'$ number of papers. 

Conversely, fits to model 1 actually seem to converge to a sharp prediction as more data is used, as illustrated by the dashed and dot-dashed curves using 50 and 70 data points respectively (which are virtually on top of each other). The results in Fig.~\ref{fig:fitsm1m2} indicate that model 1 has better forecasting features than model 2 and is hence more useful in estimating the future $N(T)$ based on the current data.

\begin{figure}[t]
\includegraphics[width=3.2in]{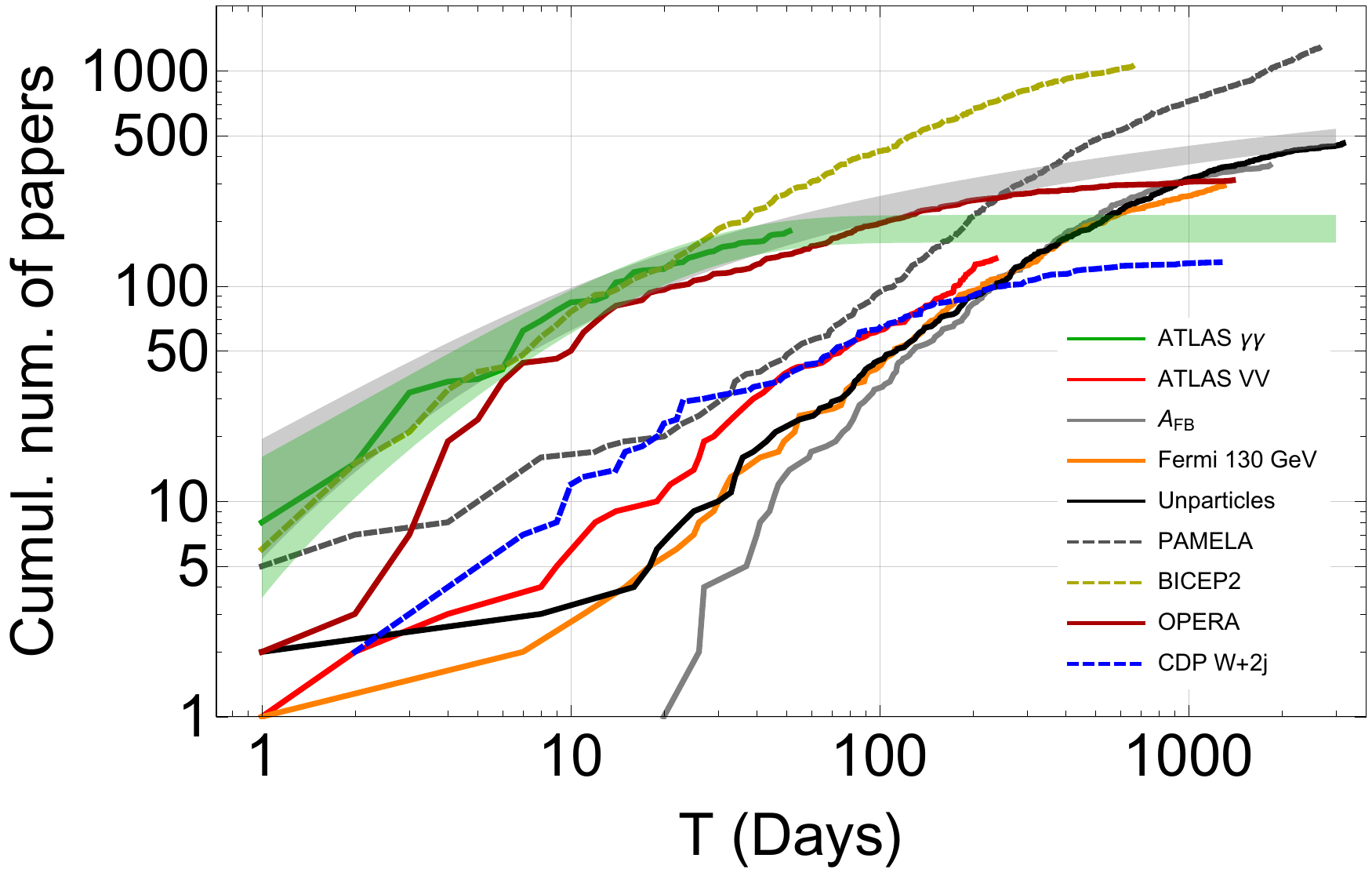}
\includegraphics[width=3.2in]{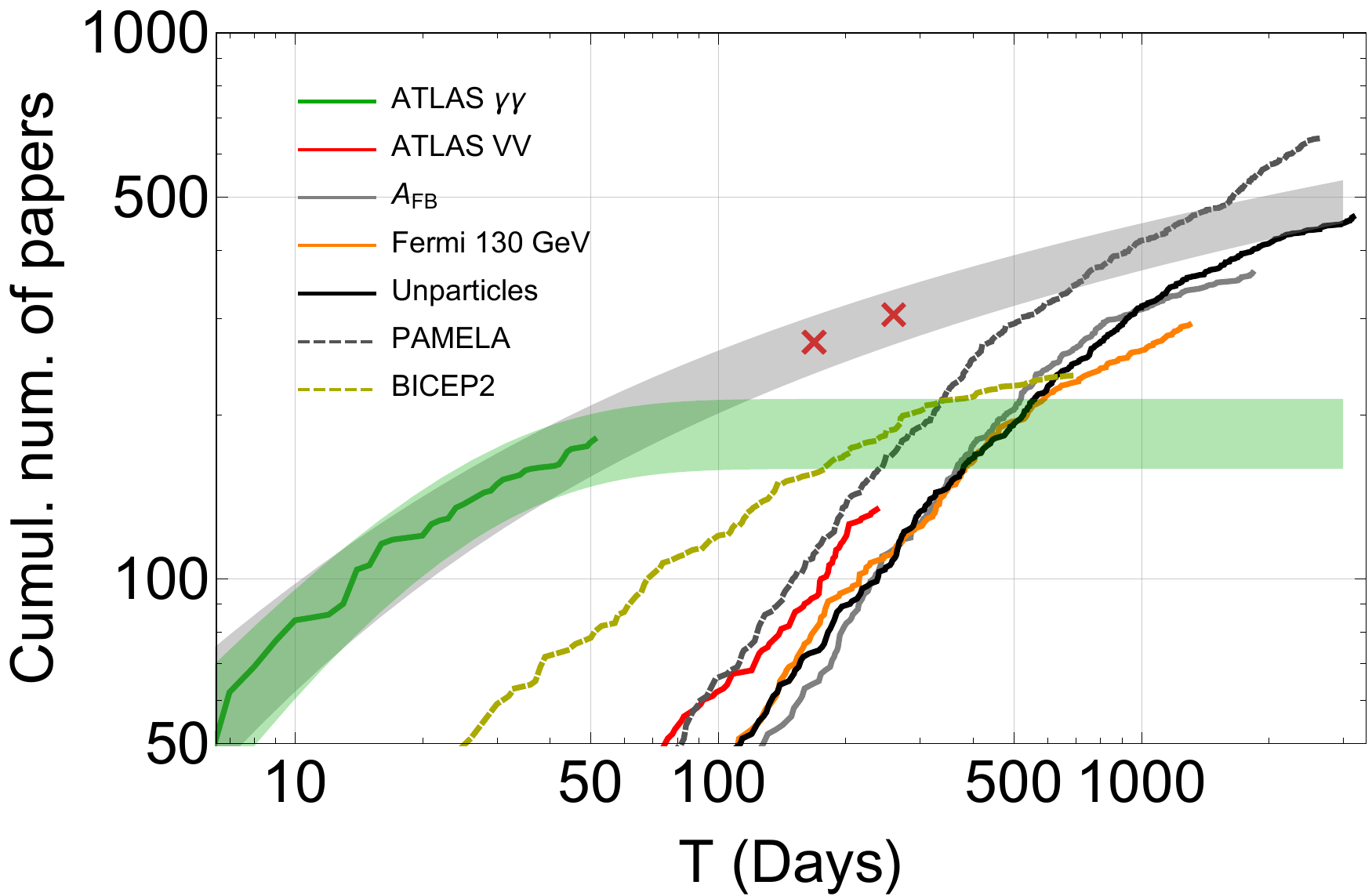}
\caption{Top panel: summary of $N(T)$ curves for 9 instances of ambulance chasing. The gray(green)  band shows the current fit to $N(T)$ in case of the ATLAS di-photon result for model 1(2), and illustrates the future projection of $N(T)$. Lower panel: The same as the top panel, but emphasizing the region where the data curves converge. The red crosses represent the most likely value for the di-photon $N(T)$  on Jun. 1. and Sep. 1. 2016. for model 1. The dashed curves represent PAMELA and BICEP 2 data, taking into account only the preprints with the hep-ph arXiv suffix in order to limit the data to the contributions from the particle physics community only.} \label{fig:alldata} 
\end{figure}

Plotting the data sets together reveals more interesting features of $N(T)$. Fig.~\ref{fig:alldata} shows the results. It is fascinating to notice that with the exception of PAMELA, BICEP2 and CDF $W+2j$, all $N(T)$ curves tend to converge to a similar functional form as $t\rightarrow \infty$. The OPERA result seems to converge to the other curves, but this is likely a coincidence, as the theoretical prediction reached the cut-off about 200 days after the announcement. Even though the PAMELA and BICEP 2 data do not match the pattern, the two sets also seem to be converging to a common functional form of their own.   At the moment, I have no clear explanation of the convergence, but suspect that it has to do something with the total number of people in the particle physics community. This could, at least in part, explain the discrepancy between the collider based results and astro-particle results, as the astro-particle results are subject to a larger audience (including particle, astro-physics and possibly cosmology communities) and should hence result in a larger overall number of publications.  Indeed, the lower panel of Fig.~\ref{fig:alldata} shows the PAMELA and BICEP 2 data taking into account only the preprints with the \verb|[hep-ph]| label in order to isolate the contributions from the particle physics community only. The sub-set of data shows a better agreement with the overall pattern of convergence at large times. 

Fig.~\ref{fig:alldata} also shows the projection for the most recent instance of ambulance chasing as the gray $2\sigma$ band for the ATLAS di-photon data fit. If model 1 is accurate, the total number of publications on the ATLAS di-photon result should not be able to exceed 310(340) papers by the beginning(end) of Summer 2016, based on the data from the first 50 days since Dec. 15. 2015. More specifically, the model predicts that the number of papers on the di-photon resonance on Jun 1. and Sep. 1. 2016 should be \footnote{Live updates of ATLAS di-photon $N(T)$ and the predictions for $N(T)$ at future times can be found on \url{http://cp3.irmp.ucl.ac.be/~mbackovic/}. }
\ba
	{\rm model\,\, 1:}\,\,\,\,\,\, & N_{\gamma \gamma}({\rm Jun.\, 1.\, 2016}) = 271 \pm 33\, , \nonumber \\
	{\rm model\,\, 1:}\,\,\,\,\,\, & N_{\gamma \gamma}({\rm Sep.\, 1.\, 2016}) = 304 \pm 34\, . \nonumber 
\ea

The prediction assumes that no significant announcements about the di-photon result, which could be characterized as the cut-off for the theory, appear in the mean time.  It is quite possible that a cut-off for the validity of the theory will be reached during the Summer of 2016, as the new LHC results are expected to settle the issue either during the June conferences, or at the vary least during the ICHEP conference at the end of July 2016. 

\section{Conclusions}

 In this paper I  argued that despite the complexity of ambulance chasing as a socio-scientific phenomenon,  it is possible to gain both qualitative and quantitative understanding of ambulance chasing observables. As an example, I have shown that the total number of papers in ambulance chasing, $N(T)$, displays simple scaling rules which assume that the interest in the topic and/or the number of available ideas on the topic are monotonically decaying, positive definite functions of time which go to 0  as $t\rightarrow \infty$.

Somewhat surprisingly, two parameter model fits suffice to accommodate all cases of ambulance chasing I have considered, up to some statistical fluctuations. 
The reason why the  model is able to fit the data is likely the same reason why we are able to do any calculations in physics: time evolution of the dynamical system in question is driven only by a few out of a large number of degrees of freedom (in this case the interest in the topic and the number of available ideas). 

I have analyzed the dynamics of ambulance chasing in two generic scenarios: one where the assumption is that the mean of the distribution, $\mu(t)$, which describes the number of papers which appear on any day $t$ scales as an inverse power of time, and the other where $\mu(t)$ decreases exponentially. It appears that the time evolution of $N(T)$ is on average better fit by the exponential model, while the power law model displays superior forecasting ability. 

The data on $N(T)$ seems to display more unexpected features. Out of 9 instances of ambulance chasing in particle physics I consider, 5 converge to roughly the same time evolution at large times. At this moment, this is only an empirical observation and I have no good explanation for why $N(T)$ behaves this way. The convergence of $N(T)$ to some unique curve could be related to the finite number of people in a particular field of physics which is engaged in an ambulance chasing instance. This could explain why OPERA and BICEP 2 results seem to converge to a larger overall $N(T)$, as these results are subject to communities of particle physics, astro-physics and cosmology, while the other instances are mostly collider physics results (with the exception of the Fermi result). It is also possible that the feature is simply a coincidence, but at this point I do not have enough statistics to make this conclusion. 

The model is able to predict the time evolution of $N(T)$, given an initial stream of data. The current ATLAS di-photon data suggests that the total number of papers will not exceed 310 by June 1. 2016, which I estimate should be well before the natural cut-off for the validity of the theory. More specifically, the model predicts that $N(T)$ of the ATLAS di-photon data on the date should be described by a Poisson distribution with the mean $\mu(T) =271$ papers. 

Forecasting the total number of preprints on an ambulance chasing topic could be useful to particle physics journals. The forecast  would allow journals to anticipate the load of submissions for publication and hence improve the overall effectiveness of the rejection process. 

Finally, it would be interesting to see if the same scaling rules apply in ambulance chasing instances across academia and the news, as it is possible that the same underlying assumptions about what drives the dynamics of ambulance chasing in particle physics apply elsewhere.

\textit{Acknowledgements:} I would like to thank the SNCB Belgian Railways for providing a comfortable environment on the trains where most of this work was conducted, as well as for frequent delays in the train system which provided the much needed additional time to complete the project. I'd also like to thank Andrea Giammanco and Fabio Maltoni for useful comments on the draft  and the encouragement to publish this work, as well as Michel Tytgat, Chiara Arina and Alberto Mariotti for useful discussions.

\end{document}